\documentclass[aps,prb,twocolumn,superscriptaddress]{revtex4-1}

\usepackage[dvips]{graphicx}
\usepackage{epsfig}

\usepackage[margin=2cm,head=0.5cm]{geometry}
\usepackage[latin1]{inputenc}
\usepackage{bm}
\usepackage{amsmath}
\usepackage{amssymb}
\usepackage{latexsym}
\usepackage{amsfonts}
\usepackage{epsfig}
\usepackage{color}
\usepackage[linktocpage, colorlinks=true ,linkcolor=blue, citecolor=blue]{hyperref}
\usepackage[all]{hypcap}

\newcommand{\expval}[1]{\langle #1 \rangle}

\newcommand{\nn}{\nonumber\\}

\newcommand{\bea}{\begin{eqnarray}}
\newcommand{\eea}{\end{eqnarray}}
\newcommand{\beann}{\begin{eqnarray*}}
\newcommand{\eeann}{\end{eqnarray*}}

\newcommand{\abs}[1]{{\left| #1 \right|}}

\newcommand{\ii}{\mathrm{i}}  

\begin{document}
  
\title{Broadband frequency filters with quantum dot chains}

\author{Tilmann Ehrlich}
\affiliation{Institut f\"ur Theoretische Physik, Technische Universit\"at Berlin, Hardenbergstr. 36, 10623 Berlin, Germany}
\author{Gernot Schaller}
\affiliation{Institut f\"ur Theoretische Physik, Technische Universit\"at Berlin, Hardenbergstr. 36, 10623 Berlin, Germany}
\affiliation{Helmholtz-Zentrum Dresden-Rossendorf, Bautzner Landstra{\ss}e 400, 01328 Dresden, Germany}
\email{g.schaller@hzdr.de}
\date{\today}
\begin{abstract}
Two-terminal electronic transport systems with a rectangular transmission can violate standard thermodynamic uncertainty relations.
This is possible beyond the linear response regime and for parameters that are not accessible with rate equations obeying detailed-balance.
Looser bounds originating from fluctuation theorem symmetries alone remain respected.
We demonstrate that optimal finite-sized  quantum dot chains can implement rectangular transmission functions with high accuracy and discuss the resulting violations of standard thermodynamic uncertainty relations as well as heat engine performance.
\end{abstract}
\maketitle

\section{Introduction}

In the macroscale, thermal machines have been beneficially used for centuries by now.
All of them have to obey the second law of thermodynamics~\cite{callen1985}, which practically bounds the efficiency of such machines, most prominent being the Carnot efficiency for heat engines.
In the last decades, with the on-going progress in the construction of nanoscale systems and machines~\cite{dubi2009a}, the laws of thermodynamics have been revisited from the quantum perspective~\cite{binder2019}.
Also quantum systems can be considered as heat engines by regarding their working fluid as an open quantum system~\cite{alicki1979a}.
Since for the study of open systems many established methods exist~\cite{weiss1993,mandel1995,breuer2002}, 
one can study the conversion of energies by considering e.g. alternating~\cite{kosloff2017a} or simultaneous~\cite{kosloff2014a} couplings to reservoirs 
held at different local equilibrium states.
Undoubtedly, there is significant practical relevance for thermal quantum machines, including e.g. absorption refrigerators~\cite{correa2014a} to cool electronic components or electric power generators driven by thermal gradients~\cite{esposito2009b}.
Beyond the direct applications however, also paradigmatic shifts were induced by this quest for miniaturization.

First, while for macroscale theories, only average quantities were relevant, among the notable contributions in the study of nanoscale systems was stochastic thermodynamics~\cite{seifert2012a}, where an entropy production can be associated even to individual trajectories.
This is quantified by the fluctuation theorem (FT)~\cite{crooks1999a,esposito2010c,utsumi2010a,golubev2011a} which relates the probabilities for trajectories with positive entropy production with those for the reversed trajectories with negative entropy production.
While on average this merely implies that the average entropy production is positive, the FT also bounds the ratio of currents and their fluctuations, thus having practical relevance also for the reliability of nanomachines~\cite{pietzonka2018a}.
Such thermodynamic uncertainty relations (TURs) have been universally established for rate equations obeying detailed- balance~\cite{barato2015a,pietzonka2016a,gingrich2016a}, in the linear-response regime~\cite{macieszczak2018a,guarnieri2019a}, for harmonic systems~\cite{saryal2019a}, generalized Langevin equations~\cite{dechant2018a,busiello2019a}, and for driven systems~\cite{barato2018a,koyuk2018a,koyuk2020a,vanvu2020a}.
Their predictions are accessible in simulated~\cite{marsland2019a} or real experiments~\cite{friedman2020a,pal2020a} 
and their finite-time versions~\cite{horowitz2017a} are for example useful to estimate the entropy production~\cite{manikandan2020a}.
Recently, it has been shown that looser versions of them only rely on the existence of an FT symmetry~\cite{timpanaro2019a,hasegawa2019a}.
Thus, we consider this class of uncertainty relations as most general, since it includes the other limits.
This also implies that the other bounds can be broken in regimes that they are not intended for.

Second, while in classical thermodynamics the energy content of the wall can in many circumstances be neglected, this is seldomly true for open quantum systems.
Only in the extreme weak-coupling limit one can neglect the energy contained in the interaction, such that energy changes in the reservoirs are accompanied by corresponding negative energy changes in the system and vice versa, leading to Pauli-type rate equations~\cite{schaller2014}.
Beyond the weak-coupling limit, the required methods are more involved and the proper definition of heat and work is often more subtle~\cite{campisi2011a,esposito2015b,perarnau_llobet2018a}. 
However, for stationary electronic transport a consistent thermodynamic picture can be established also beyond the weak-coupling regime.

This paper tries to address the quantum fluctuations in non-interacting electronic transport.
Specifically, we will demonstrate that in this scenario  various uncertainty relations can be violated away from linear response and detailed-balance regimes, while still respecting the ones originating from the FT symmetry.
We will for simplicity constrain ourselves to two-terminal systems operating at steady state.
We will make use of reaction-coordinate mapping techniques~\cite{martinazzo2011a}, which can equally 
well be applied to more general and also time-dependent setups. 
These allow to reorganize the reservoir into a reaction coordinate and a residual reservoir in a way that the original system is coupled only to the reaction coordinate and then the reaction coordinate is coupled to the residual reservoir.
Traditionally, this is used to shift the boundary between system and reservoir, allowing to obtain some results for e.g. the strong coupling or non-Markovian regime using weak-coupling approaches on the enlarged supersystem~\cite{strasberg2016a,newman2017a,strasberg2018a,schaller2018a}.
In contrast to this, we will use a reverted reaction-coordinate mapping~\cite{martensen2019a} to simplify the exact computation of the transmission, on which many stationary transport quantities are based.
While normally it can be obtained using e.g. nonequilibrium Greens functions~\cite{haug2008}, the sequential character of the reverse mapping allows us to perform simple optimizations.
Our essential result is that a rectangular transmission minimizes fluctuations and maximizes the thermal performance of a nano heat engine. 
We will demonstrate that finite-size quantum dot chains can be tuned to approximate rectangular transmissions and moreover that thermodynamic uncertainty relations provide suitable cost functions that can be minimized in an experiment, requiring only the measurement of current and its noise.

This article is organized as follows:
We begin with revisiting some required central quantities in Sec.~\ref{SEC:central_quantities} by discussing the Levitov-Lesovik formula and the thermodynamic uncertainty relation, which allow us to reformulate our main motivation for this paper in technical terms.
Afterwards, we explain various ways of obtaining a rectangular transmission function in quantum dot chains in Sec.~\ref{SEC:rectangular_transmission}.
We present our results on low-noise transport and maximal heat engine performance in Sec.~\ref{SEC:results}, before concluding with a summary.
In the appendix we provide details on the reverse reaction-coordinate mapping, its benchmarking with non-equilibrium Greens function results, and on heat engine performance in the ideal limit of a rectangular transmission.

\section{Electronic transport theory and central quantities}\label{SEC:central_quantities}

Transport is a genuine non-equilibrium phenomenon and in principle requires non-equilibrium thermodynamic approaches that go beyond weak-coupling or linear-response scenarios~\cite{esposito2010b}.
Even in the simplest two-terminal case, where a central system is coupled to a left and to a right reservoir, initially prepared in local thermal equilibrium states, the mere definition of currents may require some thought. 
For example, the time-dependent particle current entering the system from the left reservoir is not the same as the time-dependent particle current leaving the system to the right reservoir.
This is trivially so as the system has its own capacity to store particles.
When it comes to energy currents, one has to be more careful: 
Even for a single junction, the time-dependent energy current entering the system with Hamiltonian $H_S$ from reservoir $\nu$
-- defined by the corresponding balance term in $\frac{d}{dt} \expval{H_S}$ -- is in general not the same as the time-dependent energy current leaving the reservoir $\nu$ with Hamiltonian $H_B^\nu$ -- defined by $-\frac{d}{dt} \expval{H_B^\nu}$.
This is the case as the interaction Hamiltonian may carry some energy as well. 
For exactly solvable systems, such time-dependent currents can for example be obtained with non-equilibrium Greens functions~\cite{haug2008}, the Feynman-Vernon influence functional approach~\cite{jin2010a,yang2014b}, or by simply solving the Heisenberg equations of motion for quadratic operators~\cite{topp2015a,jussiau2019a}.
However, when defining such general time-dependent currents, one has to carefully specify the interface it goes through or the observable it changes.

Fortunately, things are a lot simpler in two-terminal setups that in the long-term limit evolve towards a stationary nonequilibrium state.
In this limit, the stationary currents leaving the left reservoir are the same as the stationary currents entering the right reservoir, since globally the matter and energy are conserved. 
Then, a simpler analysis is applicable that only specifies the charge and energy transfers going from left to right through the system, which we outline below.

\subsection{Levitov-Lesovik-formula}

The Levitov-Lesovik formula~\cite{levitov1993a,schoenhammer2007a} provides the long-term cumulant-generating function for independent electronic transfers through a two-terminal junction with left (L) and right (R) leads in
equilibrium described by Fermi functions $f_\nu(\omega)=[e^{\beta_\nu(\omega-\mu_\nu)}+1]^{-1}$, where $\beta_\nu$ and $\mu_\nu$ denote inverse temperature and chemical potential of lead $\nu\in\{L,R\}$, respectively.
When we consider particle transfers from left to right, it can be written as 
(unless noted, we use units with $\hbar=1$ and $k_{\rm B}=1$ throughout)
\begin{align}
C(\chi) &= \int\frac{d\omega}{2\pi} \ln \Big\{1+T(\omega) \Big[f_{LR}(\omega) (e^{+\ii\chi}-1)\nn
&\qquad+f_{RL}(\omega) (e^{-\ii\chi}-1)\Big]\Big\}\,,
\end{align}
where $0 \le T(\omega) \le 1$ denotes the transmission probability for transfers through the system at energy $\omega$ and $f_{\nu\bar\nu}(\omega) \equiv f_\nu(\omega)[1-f_{\bar\nu}(\omega)]$.
The formula above holds beyond linear-response or weak-coupling regimes. 
It captures the long-term dynamics exactly in the large-deviation sense: The contribution to cumulants of reservoir particle changes that grows linearly in time is included, but any constant contributions are missed.
The full counting statistics of particle transfers through the system is thus fully determined by a specific transmission function $T(\omega)$, and from the above formula, one is able to evaluate the cumulants of the current distribution by computing suitable derivatives with respect to the particle counting field $\chi$.
In particular, the first two derivatives yield the stationary matter current (Landauer formula~\cite{landauer1957a}) and its noise, respectively
\begin{align}\label{EQ:current_noise}
I_M &= (-\ii \partial_\chi)^1 C(\chi)|_{\chi=0}\\
&= \frac{1}{2\pi} \int T(\omega) \left[f_L(\omega)-f_R(\omega)\right] d\omega\,,\nn
S_M &= (-\ii \partial_\chi)^2 C(\chi)|_{\chi=0}\nn
&= \frac{1}{2\pi} \int T(\omega) \left[f_{LL}(\omega)+f_{RR}(\omega)\right] d\omega\nn
&\qquad+\frac{1}{2\pi} \int T(\omega)[1-T(\omega)] \left[f_L(\omega)-f_R(\omega)\right]^2 d\omega\,.\nonumber
\end{align}
The current above is the exact stationary limit of the time derivative of reservoir particle number operators $I_M = -\lim_{t\to\infty} \frac{d}{dt} \expval{N_B^L} = + \lim_{t\to\infty} \frac{d}{dt} \expval{N_B^R}$.

Since the transmission describes ballistic energy transfers at energy $\omega$, the cumulant-generating function can be 
straightforwardly extended to a version with an energy counting field $\xi$
\begin{align}\label{EQ:cgf}
C(\chi,\xi) &= \int\frac{d\omega}{2\pi} \ln \Big\{1+T(\omega) \Big[f_{LR}(\omega) (e^{+\ii(\chi+\omega\xi)}-1)\nn
&\qquad+f_{RL}(\omega) (e^{-\ii(\chi+\omega\xi)}-1)\Big]\Big\}\,,
\end{align}
from which also the energy current from left to right
\begin{align}\label{EQ:energy_current}
I_E&=(-\ii \partial_\xi)^1 C(\chi,\xi)|_{\chi=\xi=0}\nn
&=\frac{1}{2\pi} \int \omega T(\omega) \left[f_L(\omega)-f_R(\omega)\right] d\omega
\end{align}
and its noise can be obtained analogously by performing derivatives.
Here, the energy current above can also be expressed as exact steady-state limit of the reservoir energy changes $I_E = -\lim_{t\to\infty} \frac{d}{dt} \expval{H_B^L} = + \lim_{t\to\infty} \frac{d}{dt} \expval{H_B^R}$.

We note that the term in square brackets in~\eqref{EQ:cgf} and thus also the generalized cumulant-generating function obeys the fluctuation theorem symmetry
\begin{align}\label{EQ:ft_symmetry}
C(\chi,\xi) = C(-\chi + \ii (\beta_L \mu_L - \beta_R \mu_R),-\xi+\ii (\beta_R-\beta_L))\,,
\end{align}
which implies the (long-term) fluctuation theorem~\cite{crooks1999a,andrieux2006a,esposito2009a} for the probability of observing trajectories with $n$ particles and total energy $E$ transferred from left to right
\begin{align}
\lim_{t\to\infty} \frac{P_{+n,+E}(t)}{P_{-n,-E}(t)} = e^{n (\beta_L \mu_L - \beta_R \mu_R) + E(\beta_R-\beta_L)}\,.
\end{align}
The term in the exponent approximates in the long-term limit the (total) entropy production $\Sigma$ of such trajectories.
In this limit, for systems admitting stationary currents, both $\expval{n}_t$ and $\expval{E}_t$ will rise linearly in time, such that the exponent captures the (dominant) long-term contribution to the entropy production by the reservoirs, but misses the (finite) contribution by the system and also any finite contributions by the reservoirs.
The above formula implies that this dominant contribution to the average entropy production $\expval{\Sigma}$ is always positive.
Additionally, it has been shown~\cite{nenciu2007a,topp2015a} that the associated long-term entropy production rate is also positive
\begin{align}\label{EQ:entprodrate}
\sigma = I_M (\beta_L \mu_L - \beta_R \mu_R) + I_E (\beta_R-\beta_L) \ge 0\,,
\end{align}
where the currents relate to the expectation values of the stochastic variables via $I_M = \lim\limits_{t\to\infty} \frac{\expval{n}_t}{t}$ and $I_E = \lim\limits_{t\to\infty} \frac{\expval{E}_t}{t}$.
Notably, these relations hold without any prior assumption on the system-reservoir coupling strength.

\subsection{Thermodynamic uncertainty relations}

TURs are a consequence of the second law of thermodynamics in presence of multiple reservoirs.
They relate the fluctuations and average values of stochastic quantities (currents) with the overall entropy production (rate).
For systems coupled to Markovian reservoirs satisfying detailed-balance it has been generally shown~\cite{barato2015a,gingrich2016a,pietzonka2016a,horowitz2017a} that
$[\expval{j_i^2}-\expval{j_i}^2]/\expval{j_i}^2 \ge 2/\sigma$, where the entropy production rate $\sigma=\sum_i j_i A_i$ is decomposed into fluxes $j_i$ 
and corresponding affinities $A_i$ as in Eq.~(\ref{EQ:entprodrate}).
For heat engines, such a bound imposes limits e.g. on their efficiency~\cite{pietzonka2018a}.
To simplify this standard TUR (STUR) a bit, we consider a system with two terminals held at equal temperatures $\beta_L=\beta_R=\beta$ and chemical potential difference $V=\mu_L-\mu_R$.
Then, the stationary entropy production rate~\eqref{EQ:entprodrate} is proportional to the matter current $\sigma = \beta I_M V$, and the STUR inequality reads~\cite{polettini2016a}
\begin{align}\label{eq:stur}
\beta V \frac{S_M}{I_M} \ge 2\,.
\end{align}
It can be rigorously proven that this relation holds for Markovian rate equations satisfying detailed-balance~\cite{gingrich2016a} but also for harmonic systems~\cite{saryal2019a}.
It can be broken for parameter regimes that do not admit a rate-equation description with detailed-balance, which e.g. happens in stationary electronic transport setups~\cite{agarwalla2018a,martensen2019a}, for systems subject to feedback loops~\cite{potts2019a}, and for driven systems~\cite{cangemi2020a}.

In the linear response regime, a looser bound has been derived also for quantum systems that are not necessarily subject to a Markovian evolution~\cite{guarnieri2019a}.
For the isothermal two-terminal setup the linear-response TUR (LTUR) inequality reads
\begin{align}\label{eq:ltur}
\beta V \frac{S_M}{I_M} \ge 1\,.
\end{align}

Only based on a fluctuation theorem symmetry~(\ref{EQ:ft_symmetry}) one can derive fluctuation theorem TURs (FTURs) such as~\cite{timpanaro2019a,hasegawa2019a}
\begin{align}
\frac{\expval{Q_i^2}-\expval{Q_i}^2}{\expval{Q_i}^2} 
\ge \frac{1}{\sinh^2\left[g\left(\frac{\expval{\Sigma}}{2}\right)\right]}
\ge \frac{2}{e^{\expval{\Sigma}}-1}\,,
\end{align} 
where $Q_i$ is an integrated current (such as transferred particle number) and $\Sigma$ the total entropy production.
In the first inequality the function $g(x)$ is defined implicitly by $g(x) \tanh(g(x)) = x$.
Thus, for small $x$ we can write $g(x) \approx \sqrt{x}$ whereas for large $x$ we have $g(x) \approx x$.
Now, considering the particle transfers with $Q_i=n$, we have in the long-term limit (neglecting constant contributions) $\expval{Q_i} \approx I_M t$, $\expval{Q_i^2}-\expval{Q_i}^2 \approx S_M t$, and in particular the entropy production rises linearly in time $\expval{\Sigma} \approx \sigma t = I_M \beta V t$.
Inserting this in the above equation, we thus see that both expressions on the r.h.s. tend to zero faster than $1/t$.
Multiplying the above by $I_M \beta V t$ and performing the limit $t\to\infty$ we obtain a trivial bound on the ratio of noise and current (or the Fano factor $F\equiv S_M/\abs{I_M}$)
\begin{align}\label{eq:ftur}
\beta V \frac{S_M}{I_M} \ge 0\,.
\end{align}
Thus, one may conjecture that STUR and LTUR relations~\eqref{eq:stur} and~\eqref{eq:ltur} can be broken for systems that do not obey simple rate equations and beyond the linear response regime.

\subsection{Motivation: Minimizing uncertainty}

It is in fact quite simple to see that for a box-shaped rectangular transmission~\cite{martensen2019a}
\begin{align}
T(\omega) = \Theta(\omega-\omega_{\rm min}) \Theta(\omega_{\rm max}-\omega)
\end{align}
that allows perfect energy transfers in the transmission window $[\omega_{\rm min},\omega_{\rm max}]$ and blocks transfers anywhere else,
one may reach a situation yielding a finite matter current $I_M$ with negligible noise $S_M$ (note that we only discuss the contributions rising linearly in time and thereby neglect features such as bound states~\cite{longhi2007a,jussiau2019a} or any constant finite contributions) that saturates the FTUR bound~\eqref{eq:ftur}.
In the current and noise integrals~(\ref{EQ:current_noise}), the integration boundary will then be limited to the interval $[\omega_{\rm min},\omega_{\rm max}]$ for a rectangular transmission function. 
Now, for a sufficiently large bias voltage with the transport window enclosing the transmission window $\mu_L \gg \omega_{\rm max}$ and $\mu_R \ll \omega_{\rm min}$ and sufficiently low temperatures $\abs{\beta_\nu \mu_\nu} \gg 1$,
we will thus have $f_L(\omega) \approx 1\qquad\forall \omega\in [\omega_{\rm min},\omega_{\rm max}]$ and likewise
$f_R(\omega) \approx 0\qquad\forall \omega\in [\omega_{\rm min},\omega_{\rm max}]$, such that
$I_M \to (\omega_{\rm max}-\omega_{\rm min})/(2\pi)$ and $S_M \to 0$ in Eq.~(\ref{EQ:current_noise}).
This clearly breaks the STUR~(\ref{eq:stur}).
Additionally, we also see that the LTUR~(\ref{eq:ltur}) is broken, which is also demonstrated by the dashed brown curve in Fig.~\ref{FIG:uncert_entrop}.
One may object that such a rectangular transmission is reached by an infinitely long and homogeneous 
chain of quantum dots (see e.g.~\cite{cha2020a} for explicit examples), and so far only violations of the
STUR bound~\eqref{eq:stur} have been demonstrated with finite quantum dot chains.
Our foundational interest in this paper is therefore to investigate whether it is possible to overcome the LTUR bound~(\ref{eq:ltur}) with a chain composed of
a finite number of dots.

\subsection{Motivation: Optimal energy filters}

In the construction of continuous heat engines~\cite{kosloff2014a}, i.e., multi-terminal open quantum systems designed to control the flow of heat in non-equilibrium environments, it may be particularly useful to filter energies e.g. via an energy-dependent spectral coupling density.
Using a broadband energy filter for example may help to construct an absorption refrigerator as follows: An electronic quantum system is weakly coupled to a source reservoir (with a high chemical potential)  via a low-energy filter.
Additionally, it is (weakly) coupled to a drain reservoir (with low chemical potential) via a high-energy filter.
If the frequency intervals of the filters do not overlap, transport will be blocked for the two-terminal setup, since the system will be filled by electrons from the source, which due to the energy filtering cannot leave through the high-energy filter to the drain.
If in contrast we couple the system additionally to a third phonon reservoir, transport becomes possible, but only by absorbing energy from the third reservoir.
Then, low-energy electrons can enter the system from the source, absorb energy from the third reservoir, and leave the system again to the drain at a higher energy, which effectively
cools the third terminal by investing chemical work.

In this paper however we focus on two-terminal systems, where rectangular transmission functions may be used as energy filters not between system and reservoir, but between the two reservoirs.
For simplicity, we will constrain our discussion on electronic transport systems with energy and matter conservation, 
where matter and energy currents from left to right are given by the Landauer 
formulas~(\ref{EQ:current_noise}) and~(\ref{EQ:energy_current}), respectively.
From this, we can construct the heat currents leaving the left or right reservoirs via 
$J_L = I_E-\mu_L I_M$ and $J_R = -(I_E-\mu_R I_M)$.
Applying only a potential bias would drive  particles from high chemical potential to low chemical potential.
Likewise, a simple temperature bias would drive a heat flow from hot to cold reservoir.
Interesting dynamics may however arise in the case where thermal and potential gradients are tilted.
Without loss of generality we consider here the case of a cold left reservoir at high chemical potential and a hot right reservoir at lower chemical potential
\begin{align}
\beta_L > \beta_R\,,\qquad
\mu_L > \mu_R\,.
\end{align}

\begin{figure}
\includegraphics[width=0.45\textwidth,clip=true]{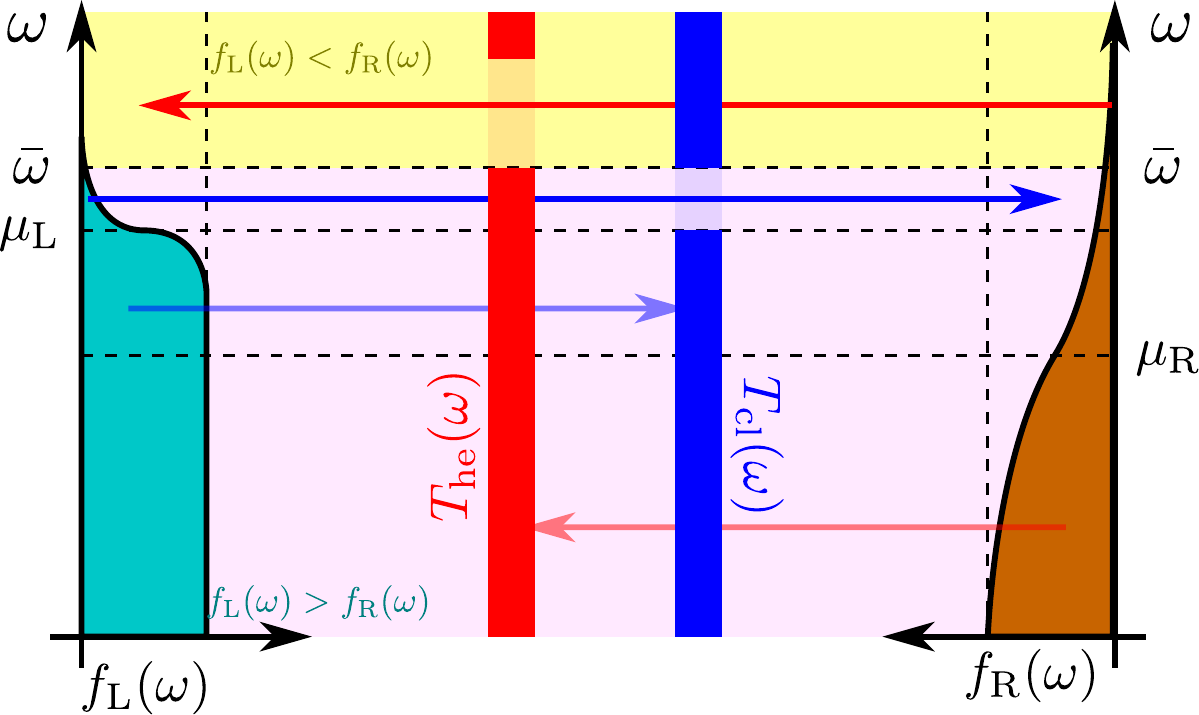}
\caption{\label{FIG:heatengine_refrigerator}
Energy sketch of left and right Fermi functions for $\beta_L>\beta_R$ and $\mu_L>\mu_R$.
Both Fermi functions are equal at $\bar\omega=\frac{\beta_L\mu_L-\beta_R\mu_R}{\beta_L-\beta_R}$.
Above this threshold, we have $f_L(\omega)<f_R(\omega)$, and below we have $f_L(\omega)>f_R(\omega)$.
By appropriately placing a rectangular transmission window as depicted, we can achieve optimal heat engine (red transmission) or refrigerator performance (blue transmission) at fixed thermal parameters.
To maximize the electric power, the transmission window should range in $[\bar\omega,\infty]$, which can be approximated with a wide rectangular transmission (red).
To maximize the cooling energy current, the transmission window should range in $[\mu_L,\bar\omega]$.
Deviations from the perfect rectangular shape and position will then reduce performance.
}
\end{figure}

Then, the cold reservoir may be cooled by investing chemical work and suitably positioning the transmission window.
As one can see from constructing the heat current via~\eqref{EQ:energy_current} and~\eqref{EQ:current_noise}, the positive contributions to the cooling current are maximized when the transmission covers the interval where $(\omega-\mu_L)[f_L(\omega)-f_R(\omega)]>0$.
Similarly, one may generate chemical work (in this case electric power by driving electrons against the bias) by using heat from the hot (right) reservoir as is depicted in Fig.~\ref{FIG:heatengine_refrigerator}.
To understand the direction of matter and energy flows, it is sufficient to realize that at one particular energy
\begin{align}
\bar\omega =\frac{\beta_L\mu_L-\beta_R\mu_R}{\beta_L-\beta_R} = \mu_L + \frac{\beta_R}{\beta_L-\beta_R} (\mu_L-\mu_R)
\end{align}
the two Fermi functions are equal $f_L(\bar\omega)=f_R(\bar\omega)$.
Hence, with a rectangular transmission we can select energy intervals to control the flow of heat.

In appendix~\ref{SEC:cooling_rect} and~\ref{SEC:heatengine_rect} we explore how the maximum heat current and the associated coefficient of performance for cooling the cold (left) reservoir by investing chemical work and similarly the power and efficiency of generating electric power from heat of the hot (right) reservoir behave in case of a perfectly rectangular transmission.
Our second -- more applied -- motivation in this paper is therefore how thermodynamic performance of a device with a nearly rectangular transmission is improved. 

\section{Generating rectangular transmissions}\label{SEC:rectangular_transmission}

\subsection{Transmission formula for a single dot}

The transmission of a single quantum dot that is coupled via general energy-dependent tunnel rates (spectral coupling densities) $\Gamma_\nu(\omega)$ to two leads is given by~\cite{haug2008,topp2015a}
\begin{align}\label{EQ:transmission_formula}
T(\omega) &= \frac{\Gamma_L(\omega)\Gamma_R(\omega)}{\left[\omega-\epsilon-\Lambda(\omega)\right]^2+\left[\frac{\Gamma_L(\omega)+\Gamma_R(\omega)}{2}\right]^2}\,,\nn
\Lambda(\omega) &\equiv \frac{1}{2\pi} {\cal P} \int \frac{\Gamma_L(\omega')+\Gamma_R(\omega')}{\omega-\omega'} d\omega'\,.
\end{align}
The functions $\Gamma_\nu(\omega)$ can be defined microscopically: For a multi-site system, where the $i$-th site (with annihilation/creation operators $d_i$/$d_i^\dagger$) is coupled by a tunnel Hamiltonian $H_{\rm tunnel}=\left(d_i \sum_k t_{k\nu} c_{k\nu}^\dagger + {\rm h.c.}\right)$ to a fermionic reservoir $H_{\rm res} = \sum_k \epsilon_{k\nu} c_{k\nu}^\dagger c_{k\nu}$, they are given by $\Gamma_\nu(\omega) = 2\pi \sum_k \abs{t_{k\nu}}^2 \delta(\omega-\epsilon_{k\nu})$.

From the above formula, a way to generate a rectangular transmission is to consider a universe Hamiltonian given by an infinitely long and homogeneous chain of quantum dots 
$H = \epsilon \sum_i d_i^\dagger d_i + T \sum_i [d_i^\dagger d_{i+1} + d_{i+1}^\dagger d_i]$.
Considering one of these qantum dots as the system, this leads to identical semicircular spectral coupling densities $\Gamma_\nu(\omega) = 2T \sqrt{1-(\omega-\epsilon)^2/(4 T^2)}$,  from 
which one would obtain a perfectly rectangular transmission function with $\omega_{\rm min} = \epsilon-2T$ and $\omega_{\rm max}=\epsilon+2T$.
This however would require experimental control over an infinitely large number of degrees of freedom, which appears unrealistic.
Therefore, in this paper, we will address the question whether it is possible to achieve sharp frequency filters by using a finite number of fine-tuned quantum dots that are coupled
to reservoirs characterized by a structureless (flat) spectral coupling density.

\subsection{Mapping relation}

In principle, the transmission of a chain of quantum dots can be obtained by nonequilibrium Greens function techniques~\cite{meir1992a,meir1993a,haug2008,wang2014a}.
This requires some matrix inversions (which for longer chains can only be performed numerically) and also the knowledge of the free Greens function, which is e.g. known for infinitely-long tight-binding chains~\cite{economou2006}.

\begin{figure}[ht]
\includegraphics[width=0.45\textwidth,clip=true]{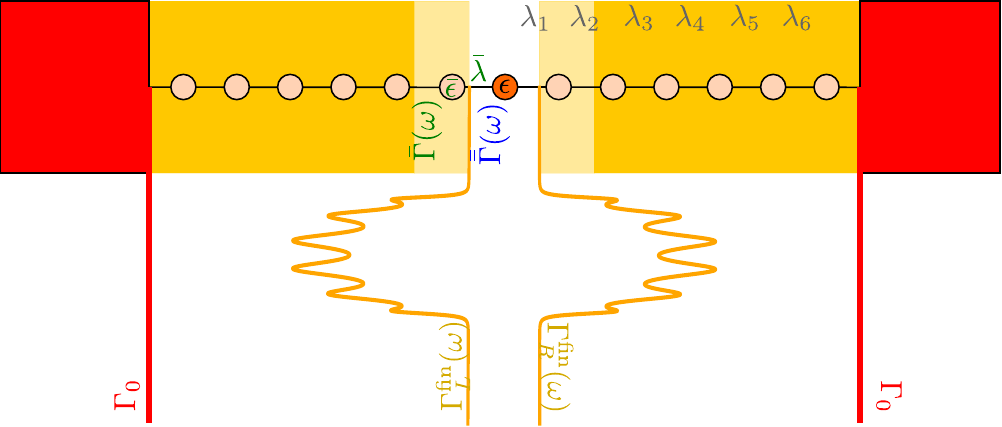}
\caption{\label{FIG:quantum_dot_chain}
Sketch of the mapping procedure: A chain with varying (but symmetric) tunnel amplitudes $\lambda_n$ and coupled to wideband reservoirs (with constant spectral coupling density $\Gamma_\nu(\omega)=\Gamma_0$, red) is mapped by sequential reverse reaction coordinate mappings to a single quantum dot coupled to two highly structured reservoirs
(with energy-dependent spectral coupling density $\Gamma_\nu^{\rm fin}(\omega)$, orange).
Blue and green symbols illustrate Eq.~\eqref{EQ:mapping} when used to effectively transfer the system-reservoir boundary (dark and light orange) by one dot.
To approach a rectangular transmission function, we adjust accessible parameters (such as the inner-most tunnel amplitude $\bar\lambda$) to minimize a suitable cost function at each step.
With the resulting optimal spectral coupling density, we insert two new dots left and right to the central one and optimize again, such that a chain with 13 dots (parameters given in Tab.~\ref{TAB:optparam}) and wideband reservoirs has the same transmission as a single-dot with highly structured reservoirs.
Intermediate spectral coupling densities are exposed in App.~\ref{APP:benchmark}.}
\end{figure}

Conventionally, the coupling between system and reservoir assumes the form of a star connecting a mode of the system (the outer dots of the chain) with all modes of the reservoir.
Such configurations generally arise as intermediate configurations of Bogoliubov-transformed system-reservoir scenarios~\cite{woods2014a,nazir2019a}.
Knowing that we can directly compute the transmission for a single dot via~\eqref{EQ:transmission_formula}, we may as well revert such schemes and 
successively map a chain coupled to wideband reservoirs (i.e., with constant tunnel rates) to a single quantum dot that is coupled to highly-structured reservoirs,
see Fig.~\ref{FIG:quantum_dot_chain}.
For a chain where the dot at the end has on-site energy $\bar\epsilon$ and is tunnel-coupled to its neighbour in the chain via amplitude $\bar\lambda$ and additionally to 
its reservoir $\nu$ via the energy-dependent tunnel rate (spectral coupling density) $\bar{\Gamma}_\nu(\omega)$ (central part of the figure), 
we can perform an inverse reaction-coordinate mapping~\cite{martensen2019a} yielding the spectral coupling density of the chain shortened by the external dot
\begin{align}\label{EQ:mapping}
\bar{\bar{\Gamma}}_\nu(\omega) = \frac{\bar{\lambda}^2 \bar{\Gamma}_\nu(\omega)}{\left(\omega-\bar{\epsilon}-{\cal P} \int \frac{\bar{\Gamma}_\nu(\omega')}{\omega-\omega'} \frac{d\omega'}{2\pi}\right)^2
+ \left(\frac{\bar{\Gamma}_\nu(\omega)}{2}\right)^2}\,.
\end{align}
Spectral coupling densities obey a scaling relation:
If the global Hamiltonian (system, interaction and reservoir) is scaled by a constant $\alpha$, this also scales the spectral coupling density by $\alpha$.
As a sanity check, we note that the mapping above preserves this scaling property.
Applying the mapping recursively to both ends of the chain, this will eventually lead to a single dot remaining for which we can directly apply the transmission formula~\eqref{EQ:transmission_formula} .
While at first the evaluation of the principal-value integral in the above equation may seem challenging, we note that this can be at least partially performed analytically.
Obvious examples are flat spectral coupling densities $\bar{\Gamma}_\nu(\omega) = \Gamma_0$, for which the first mapping just yields a Lorentzian function characterized by two poles.
If the original spectral coupling density has $L$ known poles $z_i^n$ in the upper complex half-plane
\begin{align}\label{EQ:specdens_rep}
\bar{\Gamma}_\nu(\omega) = \frac{\bar{\gamma}_\nu}{\prod_{j=1}^L (\omega-\bar{z}_j) (\omega-\bar{z}_j^*)}\,,
\end{align}
then the transformed spectral coupling density will have $(L+1)$ poles $\bar{\bar{z}}_i$ in the upper complex half-plane, which can be found numerically~\cite{martensen2019a} as detailed in App.~\ref{APP:mapping}.
The formula for the transmission~\eqref{EQ:transmission_formula} is formally equivalent to the transformation~\eqref{EQ:mapping}, such that if $\Gamma_L(\omega)$ and $\Gamma_R(\omega)$ are both of the form~(\ref{EQ:specdens_rep}) with $L$ and $R$ poles, respectively, then also the transmission resulting from this formula can be written as
$T(\omega) = \frac{t_0}{\prod_{j=1}^{L+R+1} (\omega-\tilde{z}_j) (\omega-\tilde{z}_j^*)}$ with $L+R+1$ complex conjugate pole pairs $\tilde{z}_j$ that can be found numerically as well.

\subsection{Optimization procedure}

Various optimization schemes are conceivable. 
For example, considering a target (Tg) rectangular transmission with $\omega_{\rm min} = \epsilon_{\rm Tg}-2 T_{\rm Tg}$ and $\omega_{\rm max} = \epsilon_{\rm Tg} + 2 T_{\rm Tg}$, we may numerically minimize a cost function 
\begin{align}\label{EQ:cost1}
C_1(\{\epsilon_n\},\{\lambda_n\}) = \int \left[T_{\rm Tg}(\omega) - T(\{\epsilon_n\},\{\lambda_n\})\right]^2 d\omega
\end{align}
with respect to all chain parameters such as on-site energies $\epsilon_n$ and tunnel amplitudes $\lambda_n$ simultaneously.
Since the transmission has to be calculated with non-equilibrium Greens function techniques~\cite{economou2006,haug2008,boehling2018a} or via nested applications of the previously described mapping procedure, such an optimization is numerically challenging due to the large number of parameters that are varied simultaneously.

Alternatively, knowing that a rectangular transmission~\eqref{EQ:transmission_formula} is generated by identical semicircular spectral coupling densities covering the same frequency interval
$\Gamma_{\rm Tg}(\omega)=2 T_{\rm tg}\sqrt{1-\left(\frac{\omega-\epsilon_{\rm Tg}}{2 T_{\rm Tg}}\right)^2} \Theta(4 T_{\rm tg}^2 - (\omega-\epsilon_{\rm Tg})^2)$, another suitable cost function   can be obtained by a distance measure between target and actual spectral coupling density instead.
One could pick the central dot of the chain as the remaining one and then optimize the remaining chain parameters to approximate the desired spectral coupling density felt by the central dot.
Such a minimization procedure would require only half the parameters, and an additional advantage would be that the resulting spectral coupling density could also serve as an energy filter for other (e.g. multi-terminal) setups. 
Trading the quality of the optimization for some numerical speedup however, we can also vary just the parameters at the system-reservoir boundary from Eq.~\eqref{EQ:mapping}
by considering the cost function
\begin{align}\label{EQ:cost2}
C_2(\bar\epsilon,\bar\lambda) = \int \left[\Gamma_{\rm Tg}(\omega)- \bar{\bar{\Gamma}}(\bar\epsilon,\bar\lambda,\omega)\right]^2 d\omega\,,
\end{align}
which we can numerically minimize with respect to $\bar\epsilon$ and $\bar\lambda$, such that the spectral coupling density of the internal dot approaches a semicircle one.
Although this works and is numerically rather efficient, it does not converge very fast. 
Additionally, the cost function above is not directly linked to observables and could not be directly followed in an experimental setup.

Therefore, we followed a slightly different procedure and considered the l.h.s. of the thermodynamic uncertainty relations~\eqref{eq:stur},~\eqref{eq:ltur}, and~\eqref{eq:ftur} instead.
We considered a symmetric triple dot chain with structured spectral coupling densities $\bar\Gamma_\nu(\omega)$ of the form~\eqref{EQ:specdens_rep} and vanishing on-site energies $\bar\epsilon=0$ throughout and minimized the experimentally accessible cost function
\begin{align}\label{EQ:cost3}
C_3(V,\bar\lambda) = \beta V \frac{S_M(V,\bar\lambda)}{I_M(V,\bar\lambda)}
\end{align}
with respect to the symmetric tunnel coupling between central and external dots $\bar\lambda$ and the voltage $V=\mu_L-\mu_R$ at some constant temperature $\beta$.
Numerically, the above cost function is evaluated by analytically computing the spectral coupling density $\bar{\bar{\Gamma}}_\nu(\omega)$ via the mapping~\eqref{EQ:mapping} and then via the derived transmission~\eqref{EQ:transmission_formula} noise and current~\eqref{EQ:current_noise}.
We kept the optimal chain parameter $\bar\lambda$ to update the spectral coupling density as $\bar{\bar{\Gamma}}_\nu(\omega) \to \bar\Gamma_\nu(\omega)$ in each iteration.
In the first step, we just considered constant spectral coupling densities $\Gamma_0$ instead and optimized $\lambda_6$ and the voltage.
In the second step, we used~\eqref{EQ:mapping} with a Lorentzian spectral coupling density and optimized $\lambda_5$ and the voltage, and so on.
That way, after six iterations of this procedure for every terminal, we obtain a low-noise device that can either be seen as a single quantum dot coupled to highly-structured reservoirs or a chain of 13 quantum dots that is coupled to wideband reservoirs at its end as sketched in Fig.~\ref{FIG:quantum_dot_chain}.
We provide the optimal tunneling amplitudes in units of the initial system reservoir coupling $\Gamma_0$ in Table~\ref{TAB:optparam}.
\begin{table}
\begin{tabular}{c||c|c|c|c|c|c}
parameter & $\lambda_6/\Gamma_0$ & $\lambda_5/\Gamma_0$ & $\lambda_4/\Gamma_0$ & $\lambda_3/\Gamma_0$ & $\lambda_2/\Gamma_0$ & $\lambda_1/\Gamma_0$\\
\hline
value & 0.5 & 0.3907 & 0.37254 & 0.36632 & 0.36348 & 0.36195
\end{tabular}
\caption{\label{TAB:optparam}
The 12 symmetric optimal tunnel amplitudes minimizing~\eqref{EQ:cost3} for a chain of 13 quantum dots from outside ($\lambda_6=\lambda_{-6}$) to inside ($\lambda_1=\lambda_{-1}$) (five digits shown).
}
\end{table}

\section{Results}\label{SEC:results}

\subsection{Breaking the isothermal (S/L)TUR}

\begin{figure}[ht]
\includegraphics[width=0.45\textwidth,clip=true]{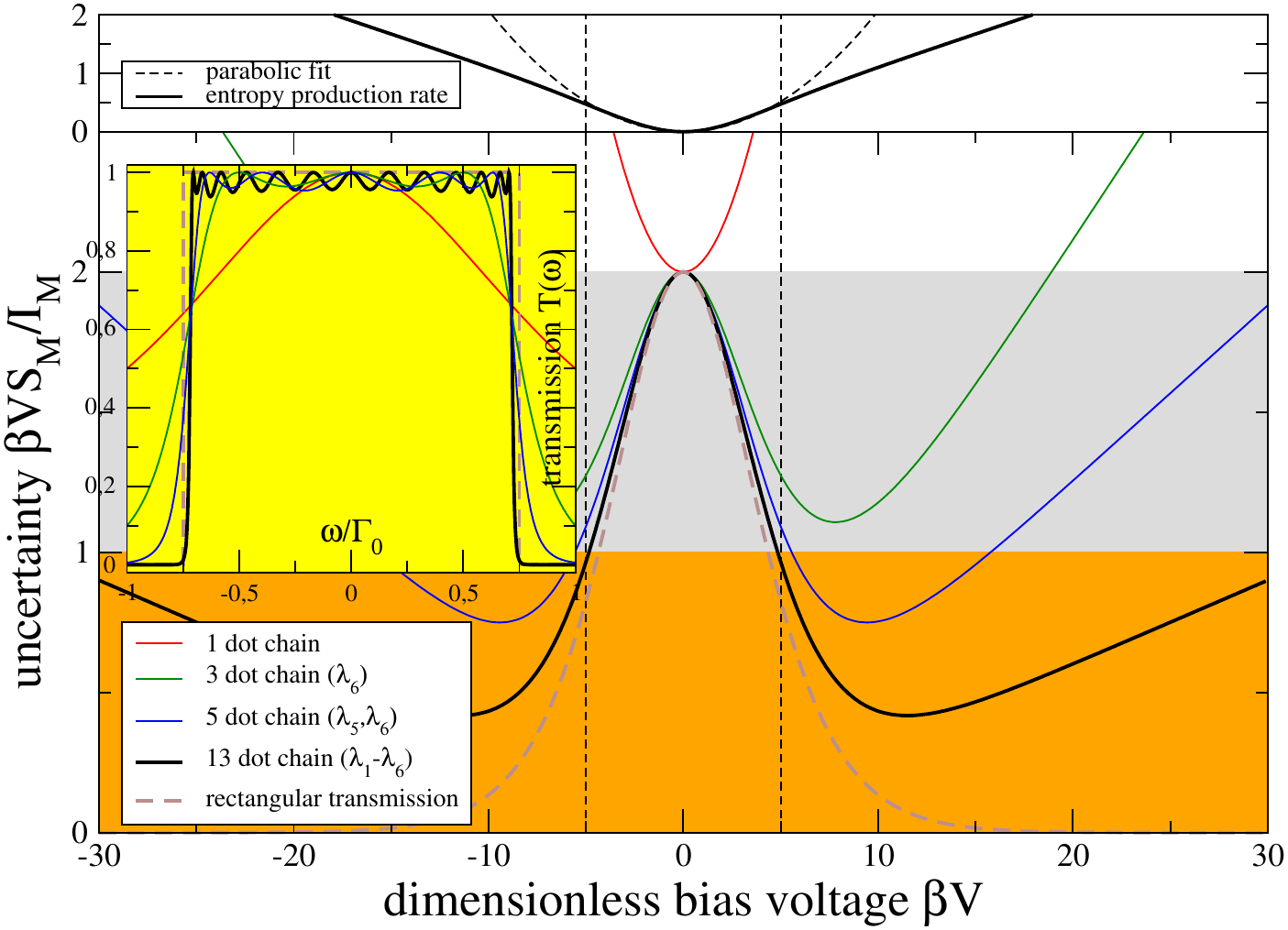}
\caption{\label{FIG:uncert_entrop}
Plot of the uncertainty quantity $\beta V S_M/I_M$ versus bias voltage for different chain lengths. 
Whereas for a single dot with wideband reservoirs (red) the STUR bound (grey region) cannot be broken, optimized triple dot chains can break it (green), 
and longer chains (blue, black) can even break the LTUR bound (orange).
This is possible in regions where the entropy production rate (top panel, 13 dot chain only) is not well approximated by a parabola, i.e., beyond the linear response regime 
(fit obtained between vertical dashed lines only). 
The inset shows the corresponding transmission functions which indeed approximate the ideal rectangular limit (dashed brown) for longer chains.
Other parameters: $\beta \Gamma_0 = 0.02$.
}
\end{figure}

With this minimization, we thus also investigate the validity of the thermodynamic uncertainty relations for our system.
We find that the current through the resulting chains violates the STUR bound~\eqref{eq:stur} already for a chain with three dots and wideband reservoirs 
(using amplitude $\lambda_6$ from Tab.~\ref{TAB:optparam}) and the LTUR bound~\eqref{eq:ltur} for a chain with five dots (improving on Ref.~\cite{martensen2019a}) 
and wideband reservoirs (using amplitudes $\lambda_5$ and $\lambda_6$ from Tab.~\ref{TAB:optparam}).
Adding further dots allows to further increase the violation as shown in Fig.~\ref{FIG:uncert_entrop}.
We see that the LTUR bound is broken beyond the nonlinear response regime, where the entropy production rate of the 13-dot chain is no longer approximated by a parabola (upper panel) and that the resulting optimized chain transmissions approach a near rectangular form (inset).
The STUR bound~\eqref{eq:stur} is always exactly matched in equilibrium, where it just reflects the Johnson-Nyquist relation for the current, see App.~\ref{APP:equilib}.
We conjecture that by improving the optimization (e.g. via optimizing the chain parameters independently for every chain length) one may achieve faster convergence to the ideal rectangular transmission limit.
We have confirmed the resulting transmissions also with an independent calculation based on non-equilibrium Greens functions~\cite{boehling2018a}, which we expose in App.~\ref{APP:benchmark}.
Additionally, we remark that optimization using the other cost functions~\eqref{EQ:cost1} and~\eqref{EQ:cost2} yielded qualitatively similar (but less drastic) violations of the STUR and LTUR
bounds (not shown).

\subsection{Maximal power and maximal cooling}

For a generic tunnel amplitude $t_{k\nu}$ and reservoir energies $\epsilon_{k\nu}$, the spectral coupling density $\Gamma_\nu(\omega) = 2\pi \sum_k \abs{t_{k\nu}}^2 \delta(\omega-\epsilon_{k\nu})$ will increase if one raises the system-reservoir coupling strength.
One might naively think that this would also always increase the currents and thereby also the power and cooling performance.
However, already when we consider a single dot with energy $\epsilon$ and wideband reservoirs described by constant spectral coupling densities $\Gamma_L(\omega)=\Gamma_R(\omega)=\Gamma_0$ (we take the parameter $\Gamma_0$ as coupling strength below), this growth of currents is only observed for small couplings $\Gamma_0$.
For this setup, the transmission~\eqref{EQ:transmission_formula} becomes a Lorentzian $T(\omega) = \Gamma_0^2/[(\omega-\epsilon)^2+\Gamma_0^2]$, and when we plot the power or cooling current -- in parameter regimes that leave them positive for small couplings $\Gamma_0$ -- as a function of the coupling strength $\Gamma_0$, we observe a turnover, such that power or cooling current decrease again beyond a certain coupling strength.
A similar turnover behaviour is found in many different setups and using various methods suitable to treat the strong-coupling regime, see e.g. Refs.~\cite{gelbwaser_klimovsky2015a,wang2015a,strasberg2018a}.
In our picture, the reason for this is that for stronger couplings the transmission windows widen (compare Fig.~\ref{FIG:heatengine_refrigerator}) and thereby unfavorable contributions to the power or cooling current arise. 
We will demonstrate that one may partially compensate for this by sharpening the transmission window.

We assume that one has a chain with a few carefully tuned quantum dots that approximate a rectangular transmission to a certain extent (compare the inset of Fig.~\ref{FIG:uncert_entrop}).
By increasing the coupling strength to the reservoirs $\Gamma_0$ (and scaling all internal parameters of the chain accordingly as in Tab.~\ref{TAB:optparam}), one stretches the width of the transmission window.
Additionally, for a given transmission window, one can vary the temperatures and chemical potentials of the reservoirs to approach the optimal situation depicted in Fig.~\ref{FIG:heatengine_refrigerator}.
We parametrize the chemical potentials as $\mu_L=\bar\mu+V/2$ and $\mu_R=\bar\mu-V/2$ and consider $V>0$ and fixed temperatures $\beta_L>\beta_R$.
Then, the frequency at which both Fermi functions coincide will vary according to  
$\bar{\omega} = \bar\mu + \frac{1}{2} \frac{\beta_L+\beta_R}{\beta_L-\beta_R} V$.
If $\bar\mu$ is chosen properly, one will extract the maximum power at some voltage.
The same holds true for a different $\bar\mu$ for the current cooling the cold reservoir.

\begin{figure}[ht]
\includegraphics[width=0.45\textwidth,clip=true]{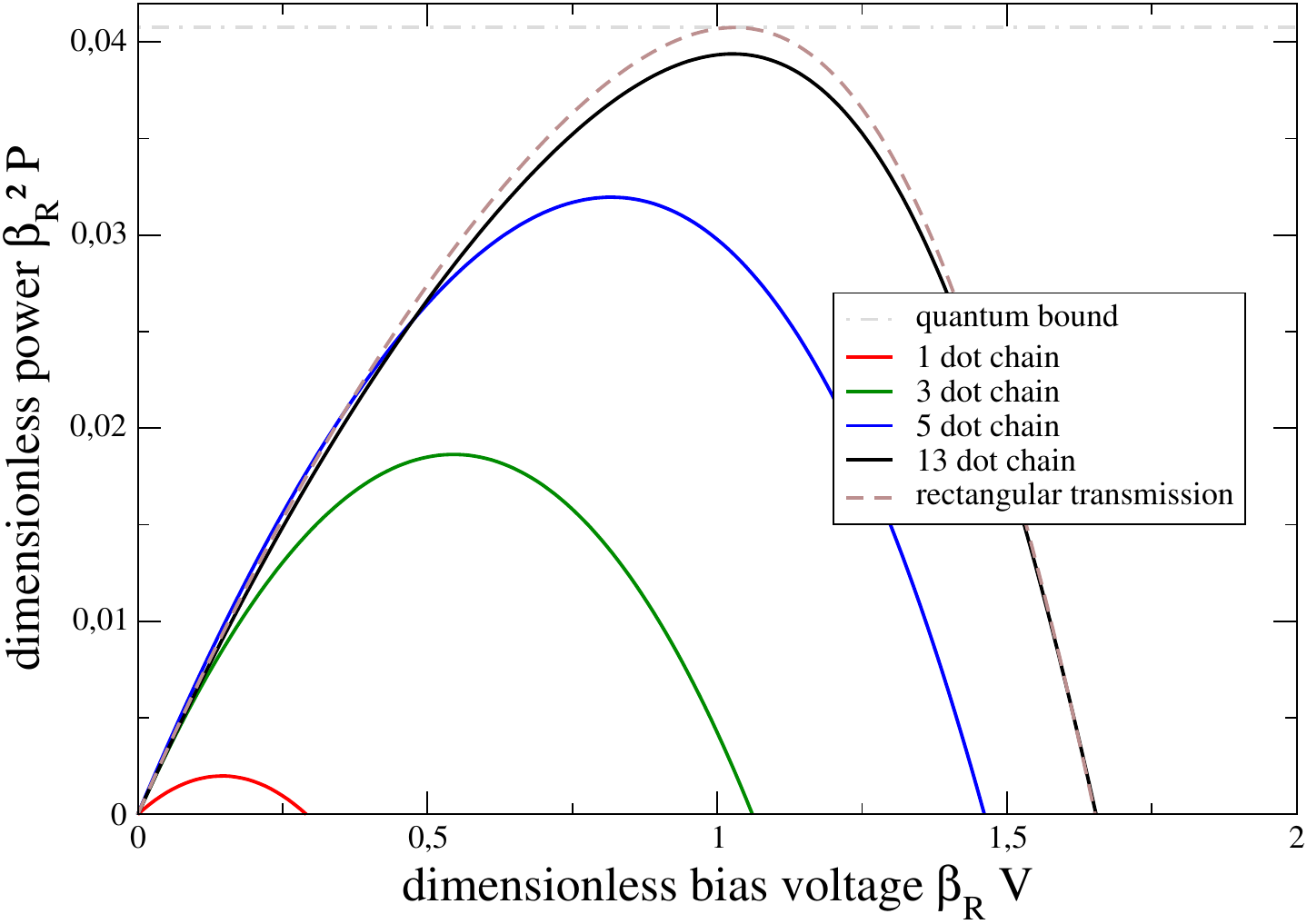}
\caption{\label{FIG:powercomp}
Plot of the generated power versus bias voltage (color coding analogous to Fig.~\ref{FIG:uncert_entrop}).
By increasing the chain length, the quality of the energy filter is improved as the transmission functions become more rectangular, nearly reaching the theoretical bound from Eq.~\eqref{EQ:maxpower}.
For comparison, the power originating from a perfectly rectangular but similar width transmission with $\beta_R\omega_{\rm min}=-3.607$ and $\beta_R\omega_{\rm max} = \beta_R \omega_{\rm min}+7.5$ is also shown (dashed brown).
Chain parameters as in Tab.~\ref{TAB:optparam}, other parameters $\beta_L=10 \beta_R$, $\beta_R \Gamma_0=5.0$, $\beta_R \bar\mu = -4.23667$.
}
\end{figure}

The power is displayed in Fig.~\ref{FIG:powercomp}.
We do now consider with $\beta_\nu \Gamma_0 \gg 1$ a strong-coupling scenario here.
Whereas in the weak-coupling regime, power increases with the coupling strength, this growth is halted and reversed beyond some coupling strength, and indeed we see that a 
single quantum dot strongly coupled to the two reservoirs produces negligible power in this regime (red curve).
The reason for this is that with increasing coupling, the energy filtering function of a single quantum dot fails.
By using fine-tuned chains as energy filters, we can compensate for this, these chains harvest significantly more power (green, blue, and black curves), nearly reaching the optimal limit for a rectangular transmission with $\omega_{\rm min} = \bar\omega$ and $\omega_{\rm max} \to \infty$, see Eq.~\eqref{EQ:maxpower}.

\begin{figure}[ht]
\includegraphics[width=0.45\textwidth,clip=true]{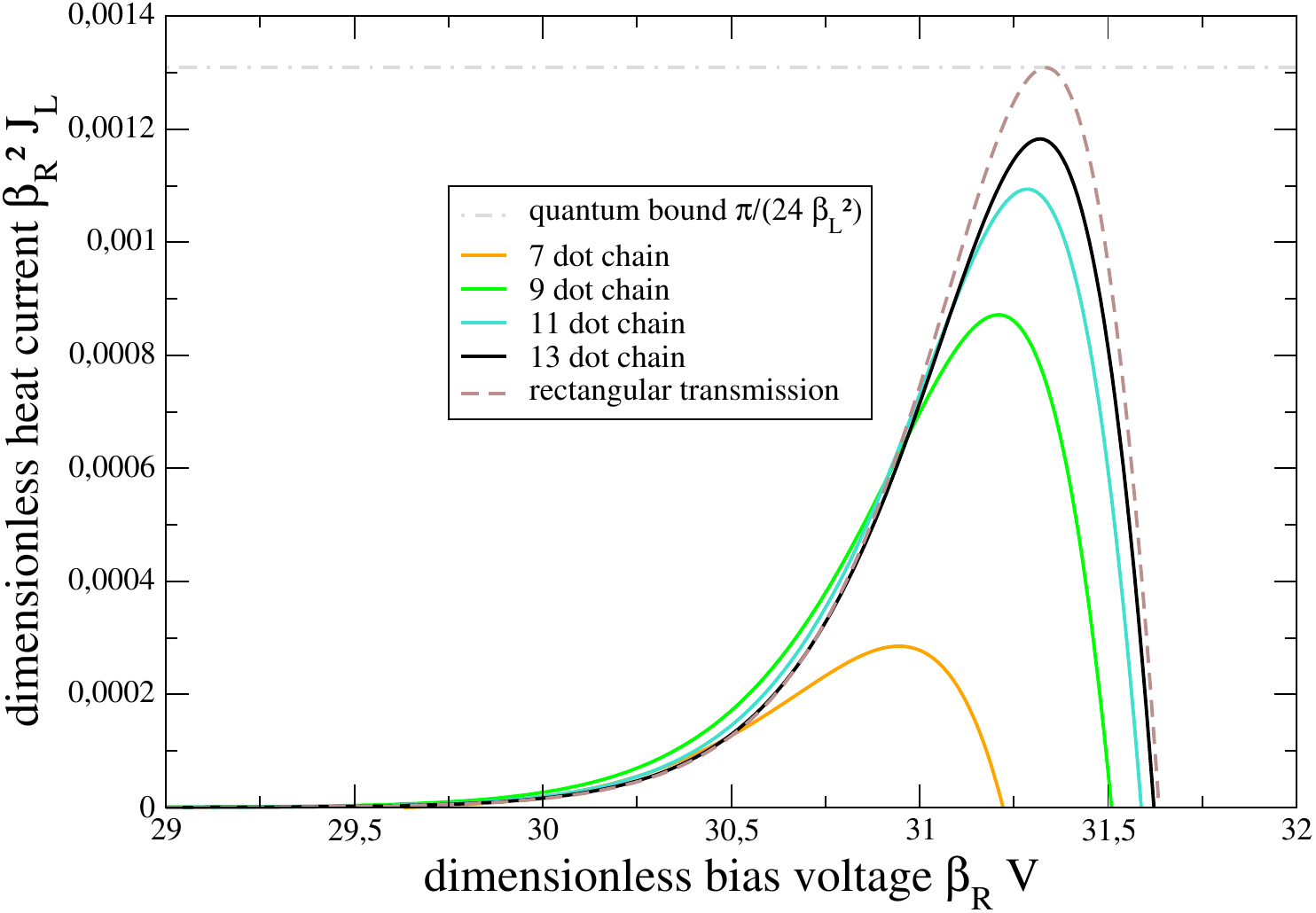}
\caption{\label{FIG:heatcurcomp}
Plot of the heat current from the left reservoir versus bias voltage.
At this coupling strength, cooling is not attainable for chains with $1$, $3$, and $5$ dots but can be achieved with longer chains.
By increasing the chain length, we nearly reach half Pendry's quantum bound from Eq.~\eqref{EQ:maxcurrent},
which is reached for the same rectangular transmission as in Fig.~\ref{FIG:powercomp} (dashed brown).
Parameters $\beta_R \bar\mu = -19.27397$, others as in Fig.~\ref{FIG:powercomp}.
}
\end{figure}

The heat current from the cold reservoir is displayed in Fig.~\ref{FIG:heatcurcomp}.
Again we see a significant improvement when the length of the chain (and thus the filtering quality) is increased.
In contrast to the extracted power, we observe that at this coupling strength, cooling function is not even attainable with chains composed of 1, 3, and 5 dots
(at smaller coupling strengths $\Gamma_0$ it would work though).
With a better filtering function, we can approach the quantum limit detailed in Eq.~\eqref{EQ:maxcurrent} for a rectangular transmission with $\omega_{\rm min} = \mu_L$ and $\omega_{\rm max}=\bar\omega$ (dashed brown).
Similar cooling performance has been observed for quantum spin hall devices~\cite{hajiloo2020a}.

\section{Summary and Conclusion}\label{SEC:summary}

We considered stationary transport of non-interacting electrons in a two-terminal setup through optimized chains of quantum dots coupled to wideband reservoirs at their ends.
Our method did not rely on weak-coupling assumptions and thus in principle allows to investigate non-Markovian and strong-coupling features.
A reverse reaction-coordinate mapping was employed to map such quantum dot chains with wideband reservoirs onto a single quantum dot coupled to structured reservoirs.
The structure of the mapping allowed to adapt the chain parameters with the goal to achieve an optimal rectangular transmission, which we then used to analyze thermodynamic uncertainty relations and thermodynamic device performance.
Admittedly, we considered the simple scenario of fully symmetric chains with vanishing on-site energies, but deviations from this can be easily taken into account with the existing method.
More challenging would be the derivation of similar mapping relations beyond onedimensional structures.

We found that the STUR and also the LTUR bounds can be broken for our setup. 
The STUR relation can be broken in a regime that is not accessible with Markovian rate equations (e.g. where the secular approximation fails). 
The LTUR relation can also be broken beyond the linear response regime, where often the entropy production rate grows slower with the bias than in the linear response regimes.
It would be interesting to investigate such relations in NEMS systems such as the electron-shuttle, where a similar reduction in the growth of entropy production can be observed~\cite{waechtler2019a}.

The thermodynamic performance of the optimized chains did approach theoretical quantum limits well.
In particular, we remark that this was achieved in a rather strong-coupling limit between system and reservoir.
In our case, the frequency filter quality of optimized chains did partially compensate for the broadening observed normally at stronger couplings, thereby opening a door to gain larger power from quantum heat engines at stronger couplings.

Thus, electronic transport setups beyond weak-coupling and linear response offer interesting options in the design of continuously operating engines.
We hope that the engineering of frequency filters by chains or other geometric configurations will find additional applications.

\acknowledgments{
G.S. gratefully acknowledges discussions with B. Agarwalla and P. Strasberg and financial support by the Helmholtz high-potential program. 
}

\bibliographystyle{unsrt}
\bibliography{references}

\begin{thebibliography}{10}

\bibitem{callen1985}
H.~B. Callen.
\newblock {\em Thermodynamics and an Introduction to Thermostatistics}.
\newblock John Wiley and Sons, 1985.

\bibitem{dubi2009a}
Yonatan Dubi and Massimiliano~Di Ventra.
\newblock Thermoelectric effects in nanoscale junctions.
\newblock {\em Nano Letters}, 9:97--101, 2009.

\bibitem{binder2019}
Felix Binder, Luis~A. Correa, Christian Gogolin, Janet Anders, and Gerardo
  Adesso, editors.
\newblock {\em Thermodynamics in the Quantum Regime -- Fundamental Aspects and
  New Directions}, volume 195 of {\em Fundamental Theories of Physics}.
\newblock Springer, 2019.

\bibitem{alicki1979a}
R.~Alicki.
\newblock The quantum open system as a model of the heat engine.
\newblock {\em Journal of Physics A: Mathematical and General}, 12:L103, 1979.

\bibitem{weiss1993}
U.~Weiss.
\newblock {\em Quantum Dissipative Systems}, volume~2 of {\em Series of Modern
  Condensed Matter Physics}.
\newblock World Scientific, Singapore, 1993.

\bibitem{mandel1995}
Leonard Mandel and Emil Wolf.
\newblock {\em Optical coherence and quantum optics}.
\newblock Cambridge University Press, 1995.

\bibitem{breuer2002}
H.-P. Breuer and F.~Petruccione.
\newblock {\em The Theory of Open Quantum Systems}.
\newblock Oxford University Press, Oxford, 2002.

\bibitem{kosloff2017a}
Ronnie Kosloff and Yair Rezek.
\newblock The quantum harmonic {O}tto cycle.
\newblock {\em Entropy}, 19:136, 2017.

\bibitem{kosloff2014a}
Ronnie Kosloff and Amikam Levy.
\newblock Quantum heat engines and refrigerators: Continuous devices.
\newblock {\em Annual Review of Physical Chemistry}, 65:365, 2014.

\bibitem{correa2014a}
Luis~A. Correa, Jos\'e~P. Palao, Daniel Alonso, and Gerardo Adesso.
\newblock Quantum-enhanced absorption refrigerators.
\newblock {\em Scientific Reports}, 4:3949, 2014.

\bibitem{esposito2009b}
M.~Esposito, K.~Lindenberg, and C.~Van den Broeck.
\newblock Thermoelectric efficiency at maximum power in a quantum dot.
\newblock {\em Europhysics Letters}, 85:60010, 2009.

\bibitem{seifert2012a}
U.~Seifert.
\newblock Stochastic thermodynamics, fluctuation theorems and molecular
  machines.
\newblock {\em Reports on Progress in Physics}, 75:126001, 2012.

\bibitem{crooks1999a}
Gavin~E. Crooks.
\newblock Entropy production fluctuation theorem and the nonequilibrium work
  relation for free energy differences.
\newblock {\em Physical Review E}, 60:2721--2726, Sep 1999.

\bibitem{esposito2010c}
Massimiliano Esposito and Christian Van~den Broeck.
\newblock Three detailed fluctuation theorems.
\newblock {\em Physical Review Letters}, 104:090601, Mar 2010.

\bibitem{utsumi2010a}
Y.~Utsumi, D.~S. Golubev, M.~Marthaler, K.~Saito, T.~Fujisawa, and Gerd
  Sch\"on.
\newblock Bidirectional single-electron counting and the fluctuation theorem.
\newblock {\em Physical Review B}, 81(12):125331, 2010.

\bibitem{golubev2011a}
D.~S. Golubev, Y.~Utsumi, M.~Marthaler, and Gerd Sch\"on.
\newblock Fluctuation theorem for a double quantum dot coupled to a
  point-contact electrometer.
\newblock {\em Physical Review B}, 84(7):075323, 2011.

\bibitem{pietzonka2018a}
Patrick Pietzonka and Udo Seifert.
\newblock Universal trade-off between power, efficiency, and constancy in
  steady-state heat engines.
\newblock {\em Phys. Rev. Lett.}, 120:190602, May 2018.

\bibitem{barato2015a}
Andre~C. Barato and Udo Seifert.
\newblock Thermodynamic uncertainty relation for biomolecular processes.
\newblock {\em Physical Review Letters}, 114:158101, Apr 2015.

\bibitem{pietzonka2016a}
Patrick Pietzonka, Andre~C. Barato, and Udo Seifert.
\newblock Universal bounds on current fluctuations.
\newblock {\em Phys. Rev. E}, 93:052145, May 2016.

\bibitem{gingrich2016a}
Todd~R. Gingrich, Jordan~M. Horowitz, Nikolay Perunov, and Jeremy~L. England.
\newblock Dissipation bounds all steady-state current fluctuations.
\newblock {\em Phys. Rev. Lett.}, 116:120601, Mar 2016.

\bibitem{macieszczak2018a}
Katarzyna Macieszczak, Kay Brandner, and Juan~P. Garrahan.
\newblock Unified thermodynamic uncertainty relations in linear response.
\newblock {\em Phys. Rev. Lett.}, 121:130601, Sep 2018.

\bibitem{guarnieri2019a}
Giacomo Guarnieri, Gabriel~T. Landi, Stephen~R. Clark, and John Goold.
\newblock Thermodynamics of precision in quantum nonequilibrium steady states.
\newblock {\em Phys. Rev. Research}, 1:033021, Oct 2019.

\bibitem{saryal2019a}
Sushant Saryal, Hava~Meira Friedman, Dvira Segal, and Bijay~Kumar Agarwalla.
\newblock Thermodynamic uncertainty relation in thermal transport.
\newblock {\em Phys. Rev. E}, 100:042101, Oct 2019.

\bibitem{dechant2018a}
Andreas Dechant and Shin ichi Sasa.
\newblock Current fluctuations and transport efficiency for general {L}angevin
  systems.
\newblock {\em Journal of Statistical Mechanics: Theory and Experiment},
  2018(6):063209, jun 2018.

\bibitem{busiello2019a}
Daniel~Maria Busiello and Simone Pigolotti.
\newblock Hyperaccurate currents in stochastic thermodynamics.
\newblock {\em Phys. Rev. E}, 100:060102, Dec 2019.

\bibitem{barato2018a}
Andre~C Barato, Raphael Chetrite, Alessandra Faggionato, and Davide Gabrielli.
\newblock Bounds on current fluctuations in periodically driven systems.
\newblock {\em New Journal of Physics}, 20(10):103023, oct 2018.

\bibitem{koyuk2018a}
Timur Koyuk, Udo Seifert, and Patrick Pietzonka.
\newblock A generalization of the thermodynamic uncertainty relation to
  periodically driven systems.
\newblock {\em Journal of Physics A: Mathematical and Theoretical},
  52(2):02LT02, dec 2018.

\bibitem{koyuk2020a}
Timur Koyuk and Udo Seifert.
\newblock Thermodynamic uncertainty relation for time-dependent driving.
\newblock {\em Phys. Rev. Lett.}, 125:260604, Dec 2020.

\bibitem{vanvu2020a}
Tan Van~Vu and Yoshihiko Hasegawa.
\newblock Thermodynamic uncertainty relations under arbitrary control
  protocols.
\newblock {\em Phys. Rev. Research}, 2:013060, Jan 2020.

\bibitem{marsland2019a}
Robert Marsland, Wenping Cui, and Jordan~M. Horowitz.
\newblock The thermodynamic uncertainty relation in biochemical oscillations.
\newblock {\em Journal of The Royal Society Interface}, 16(154):20190098, 2019.

\bibitem{friedman2020a}
Hava~Meira Friedman, Bijay~K. Agarwalla, Ofir Shein-Lumbroso, Oren Tal, and
  Dvira Segal.
\newblock Thermodynamic uncertainty relation in atomic-scale quantum
  conductors.
\newblock {\em Phys. Rev. B}, 101:195423, May 2020.

\bibitem{pal2020a}
Soham Pal, Sushant Saryal, Dvira Segal, T.~S. Mahesh, and Bijay~Kumar
  Agarwalla.
\newblock Experimental study of the thermodynamic uncertainty relation.
\newblock {\em Phys. Rev. Research}, 2:022044, May 2020.

\bibitem{horowitz2017a}
Jordan~M. Horowitz and Todd~R. Gingrich.
\newblock Proof of the finite-time thermodynamic uncertainty relation for
  steady-state currents.
\newblock {\em Phys. Rev. E}, 96:020103, Aug 2017.

\bibitem{manikandan2020a}
Sreekanth~K. Manikandan, Deepak Gupta, and Supriya Krishnamurthy.
\newblock Inferring entropy production from short experiments.
\newblock {\em Phys. Rev. Lett.}, 124:120603, Mar 2020.

\bibitem{timpanaro2019a}
Andr\'e~M. Timpanaro, Giacomo Guarnieri, John Goold, and Gabriel~T. Landi.
\newblock Thermodynamic uncertainty relations from exchange fluctuation
  theorems.
\newblock {\em Phys. Rev. Lett.}, 123:090604, Aug 2019.

\bibitem{hasegawa2019a}
Yoshihiko Hasegawa and Tan Van~Vu.
\newblock Fluctuation theorem uncertainty relation.
\newblock {\em Phys. Rev. Lett.}, 123:110602, Sep 2019.

\bibitem{schaller2014}
G.~Schaller.
\newblock {\em Open Quantum Systems Far from Equilibrium}, volume 881 of {\em
  Lecture Notes in Physics}.
\newblock Springer, Cham, 2014.

\bibitem{campisi2011a}
Michele Campisi, Peter H\"anggi, and Peter Talkner.
\newblock Colloquium: Quantum fluctuation relations: Foundations and
  applications.
\newblock {\em Rev. Mod. Phys.}, 83(3):771--791, 2011.

\bibitem{esposito2015b}
Massimiliano Esposito, Maicol~A. Ochoa, and Michael Galperin.
\newblock Nature of heat in strongly coupled open quantum systems.
\newblock {\em Physical Review B}, 92:235440, Dec 2015.

\bibitem{perarnau_llobet2018a}
M.~Perarnau-Llobet, H.~Wilming, A.~Riera, R.~Gallego, and J.~Eisert.
\newblock Strong coupling corrections in quantum thermodynamics.
\newblock {\em Phys. Rev. Lett.}, 120:120602, Mar 2018.

\bibitem{martinazzo2011a}
R.~Martinazzo, B.~Vacchini, K.~H. Hughes, and I.~Burghardt.
\newblock Universal {M}arkovian reduction of {B}rownian particle dynamics.
\newblock {\em The Journal of Chemical Physics}, 134:011101, 2011.

\bibitem{strasberg2016a}
Philipp Strasberg, Gernot Schaller, Neill Lambert, and Tobias Brandes.
\newblock Nonequilibrium thermodynamics in the strong coupling and
  non-{M}arkovian regime based on a reaction coordinate mapping.
\newblock {\em New Journal of Physics}, 18:073007, 2016.

\bibitem{newman2017a}
David Newman, Florian Mintert, and Ahsan Nazir.
\newblock Performance of a quantum heat engine at strong reservoir coupling.
\newblock {\em Physical Review E}, 95:032139, 2017.

\bibitem{strasberg2018a}
Philipp Strasberg, Gernot Schaller, Thomas~L. Schmidt, and Massimiliano
  Esposito.
\newblock Fermionic reaction coordinates and their application to an autonomous
  {M}axwell demon in the strong-coupling regime.
\newblock {\em Phys. Rev. B}, 97:205405, May 2018.

\bibitem{schaller2018a}
Gernot Schaller, Javier Cerrillo, Georg Engelhardt, and Philipp Strasberg.
\newblock Electronic {M}axwell demon in the coherent strong-coupling regime.
\newblock {\em Phys. Rev. B}, 97:195104, May 2018.

\bibitem{martensen2019a}
Niklas Martensen and Gernot Schaller.
\newblock Transmission from reverse reaction coordinate mappings.
\newblock {\em European Physical Journal B}, 92:30, 2019.

\bibitem{haug2008}
H.~Haug and A.-P. Jauho.
\newblock {\em Quantum Kinetics in Transport and Optics of Semiconductors}.
\newblock Springer, 2008.

\bibitem{esposito2010b}
M.~Esposito, K.~Lindenberg, and C.~Van den Broeck.
\newblock Entropy production as correlation between system and reservoir.
\newblock {\em New Journal of Physics}, 12:013013, 2010.

\bibitem{jin2010a}
Jinshuang Jin, Matisse Wei-Yuan Tu, Wei-Min Zhang, and YiJing Yan.
\newblock Non-equilibrium quantum theory for nanodevices based on the
  {F}eynman-{V}ernon influence functional.
\newblock {\em New Journal of Physics}, 12:083013, 2010.

\bibitem{yang2014b}
Pei-Yun Yang, Chuan-Yu Lin, and Wei-Min Zhang.
\newblock Transient current-current correlations and noise spectra.
\newblock {\em Phys. Rev. B}, 89:115411, Mar 2014.

\bibitem{topp2015a}
Gabriel~E. Topp, Tobias Brandes, and Gernot Schaller.
\newblock Steady-state thermodynamics of non-interacting transport beyond weak
  coupling.
\newblock {\em Europhysics Letters}, 110:67003, 2015.

\bibitem{jussiau2019a}
\'Etienne Jussiau, Masahiro Hasegawa, and Robert~S. Whitney.
\newblock Signature of the transition to a bound state in thermoelectric
  quantum transport.
\newblock {\em Phys. Rev. B}, 100:115411, Sep 2019.

\bibitem{levitov1993a}
L.~S. Levitov and G.~B. Lesovik.
\newblock Charge distribution in quantum shot noise.
\newblock {\em JETP Letters}, 58:230, 1993.

\bibitem{schoenhammer2007a}
K.~Sch\"onhammer.
\newblock Full counting statistics for noninteracting fermions: Exact results
  and the {L}evitov-{L}esovik formula.
\newblock {\em Phys. Rev. B}, 75:205329, May 2007.

\bibitem{landauer1957a}
R.~Landauer.
\newblock Spatial variation of currents and fields due to localized scatterers
  in metallic conduction.
\newblock {\em IBM Journal of Research and Development}, 1:223, 1957.

\bibitem{andrieux2006a}
David Andrieux and Pierre Gaspard.
\newblock Fluctuation theorem for transport in mesoscopic systems.
\newblock {\em Journal of Statistical Mechanics: Theory and Experiment},
  2006:P01011, 2006.

\bibitem{esposito2009a}
M.~Esposito, U.~Harbola, and S.~Mukamel.
\newblock Nonequilibrium fluctuations, fluctuation theorems, and counting
  statistics in quantum systems.
\newblock {\em Reviews of Modern Physics}, 81:1665--1702, 2009.

\bibitem{nenciu2007a}
Gheorghe Nenciu.
\newblock Independent electron model for open quantum systems:
  {L}andauer-{B}\"uttiker formula and strict positivity of the entropy
  production.
\newblock {\em Journal of Mathematical Physics}, 48:033302, 2007.

\bibitem{polettini2016a}
Matteo Polettini, Alexandre Lazarescu, and Massimiliano Esposito.
\newblock Tightening the uncertainty principle for stochastic currents.
\newblock {\em Phys. Rev. E}, 94:052104, Nov 2016.

\bibitem{agarwalla2018a}
Bijay~Kumar Agarwalla and Dvira Segal.
\newblock Assessing the validity of the thermodynamic uncertainty relation in
  quantum systems.
\newblock {\em Phys. Rev. B}, 98:155438, Oct 2018.

\bibitem{potts2019a}
Patrick~P. Potts and Peter Samuelsson.
\newblock Thermodynamic uncertainty relations including measurement and
  feedback.
\newblock {\em Phys. Rev. E}, 100:052137, Nov 2019.

\bibitem{cangemi2020a}
L.~M. Cangemi, V.~Cataudella, G.~Benenti, M.~Sassetti, and G.~De~Filippis.
\newblock Violation of thermodynamics uncertainty relations in a periodically
  driven work-to-work converter from weak to strong dissipation.
\newblock {\em Phys. Rev. B}, 102:165418, Oct 2020.

\bibitem{longhi2007a}
S.~Longhi.
\newblock Bound states in the continuum in a single-level {F}ano-{A}nderson
  model.
\newblock {\em The European Physical Journal B}, 57:45--51, 2007.

\bibitem{cha2020a}
Moon-Hyun Cha and Jeongwoon Hwang.
\newblock Quantum transport in a chain of quantum dots with inhomogeneous size
  distribution and manifestation of 1d {A}nderson localization.
\newblock {\em Scientific Reports}, 10:16701, 2020.

\bibitem{meir1992a}
Yigal Meir and Ned~S. Wingreen.
\newblock Landauer formula for the current through an interacting electron
  region.
\newblock {\em Physical Review Letters}, 68:2512--2515, Apr 1992.

\bibitem{meir1993a}
Y.~Meir, N.~S. Wingreen, and P.~A. Lee.
\newblock Low-temperature transport through a quantum dot: The {A}nderson model
  out of equilibrium.
\newblock {\em Physical Review Letters}, 70:2601 -- 2604, 1993.

\bibitem{wang2014a}
Jian-Sheng Wang, Bijay~Kumar Agarwalla, Huanan Li, and Juzar Thingna.
\newblock Nonequilibrium {G}reens function method for quantum thermal
  transport.
\newblock {\em Frontiers of Physics}, 9:673, 2014.

\bibitem{economou2006}
E.~N. Economou.
\newblock {\em Green's functions in quantum physics}.
\newblock Springer, Berlin Heidelberg, 2006.

\bibitem{woods2014a}
M.~P. Woods, R.~Groux, A.~W. Chin, S.~F. Huelga, and M.~B. Plenio.
\newblock Mappings of open quantum systems onto chain representations and
  {M}arkovian embeddings.
\newblock {\em Journal of Mathematical Physics}, 55:032101, 2014.

\bibitem{nazir2019a}
A.~Nazir and G.~Schaller.
\newblock The reaction coordinate mapping in quantum thermodynamics.
\newblock In F.~Binder, L.~A. Correa, C.~Gogolin, J.~Anders, and G.~Adesso,
  editors, {\em Thermodynamics in the quantum regime -- Recent progress and
  outlook}, Fundamental Theories of Physics. Springer, Cham, 2019.

\bibitem{boehling2018a}
S.~B\"ohling, G.~Engelhardt, G.~Platero, and G.~Schaller.
\newblock Thermoelectric performance of topological boundary modes.
\newblock {\em Phys. Rev. B}, 98:035132, Jul 2018.

\bibitem{gelbwaser_klimovsky2015a}
David Gelbwaser-Klimovsky and Al\'an Aspuru-Guzik.
\newblock Strongly coupled quantum heat machines.
\newblock {\em The Journal of Physical Chemistry Letters}, 6:3477, 2015.

\bibitem{wang2015a}
Chen Wang, Jie Ren, and Jianshu Cao.
\newblock Nonequilibrium energy transfer at nanoscale: A unified theory from
  weak to strong coupling.
\newblock {\em Scientific Reports}, 5:11787, 2015.

\bibitem{hajiloo2020a}
Fatemeh Hajiloo, Pablo~Terr\'en Alonso, Nastaran Dashti, Liliana Arrachea, and
  Janine Splettstoesser.
\newblock Detailed study of nonlinear cooling with two-terminal configurations
  of topological edge states.
\newblock {\em Phys. Rev. B}, 102:155434, Oct 2020.

\bibitem{waechtler2019a}
Christopher~W W\"achtler, Philipp Strasberg, Sabine H~L Klapp, Gernot Schaller,
  and Christopher Jarzynski.
\newblock Stochastic thermodynamics of self-oscillations: the electron shuttle.
\newblock {\em New Journal of Physics}, 21:073009, 2019.

\bibitem{blanter2000a}
Ya.~M. Blanter and M.~B\"uttiker.
\newblock Shot noise in mesoscopic conductors.
\newblock {\em Physics Reports}, 336:1--166, 2000.

\bibitem{whitney2014a}
Robert~S. Whitney.
\newblock Most efficient quantum thermoelectric at finite power output.
\newblock {\em Physical Review Letters}, 112:130601, Apr 2014.

\bibitem{pendry1983a}
J~B Pendry.
\newblock Quantum limits to the flow of information and entropy.
\newblock {\em Journal of Physics A: Mathematical and General},
  16(10):2161--2171, jul 1983.

\bibitem{chambadal1957}
Paul Chambadal.
\newblock {\em Les Centrales Nucl\'eaires}.
\newblock Armand Colin, Paris, 1957.

\bibitem{novikov1958a}
I.I. Novikov.
\newblock The efficiency of atomic power stations (a review).
\newblock {\em Journal of Nuclear Energy (1954)}, 7(1):125--128, 1958.

\bibitem{curzon1975a}
F.~L. Curzon and B.~Ahlborn.
\newblock Efficiency of a {C}arnot engine at maximum power output.
\newblock {\em American Journal of Physics}, 43:22, 1975.

\bibitem{vandenbroeck2005a}
C.~{Van den Broeck}.
\newblock Thermodynamic efficiency at maximum power.
\newblock {\em Physical Review Letters}, 95:190602, 2005.

\end{thebibliography}

\appendix

\section{Explicit calculation of the reverse mapping}\label{APP:mapping}

Applying the reverse mapping~\eqref{EQ:mapping} once for initially flat spectral coupling densities, we obtain simple Lorentzian spectral coupling densities that can be parametrized like~\eqref{EQ:specdens_rep}.
For this class of spectral coupling densities, the reverse reaction-coordinate mapping~\eqref{EQ:mapping} can be evaluated with functional calculus (we assume first order poles only).
Using that the $z_i$ are in the upper complex plane, we can appropriately excise the singularity at $\omega$ on the real axis in the principal-value integral.
The net effect of this procedure is that it only contributes half compared to the residues of the poles $z_i$.
Absorbing the residues of the spectral coupling density in the quantities
\begin{align}
   K_j &\equiv \bar\gamma \frac{\ii}{(\bar{z}_j-\bar{z}_j^*)\prod\limits_{i\neq j} (\bar{z}_j-\bar{z}_i)(\bar{z}_j-\bar{z}_i^*)}\,, 
\end{align}
the transformed spectral coupling density becomes
\begin{align}
\bar{\bar{\Gamma}}(\omega) &= \frac{\bar\lambda^2 \bar\gamma}{\abs{(\omega-\bar\epsilon)\prod\limits_{j=1}^L (\omega-\bar{z}_j) - \sum\limits_{j=1}^L K_j \prod\limits_{i\neq j} (\omega-\bar{z}_i)}^2}\nn
&\equiv \frac{\bar{\bar{\gamma}}}{\prod\limits_{i=1}^{L+1} (\omega-\bar{\bar{z}}_i)(\omega-\bar{\bar{z}}_i^*)}\,,
\end{align}
where the poles of the transformed spectral coupling density $\bar{\bar{z}}_i$ are given by the roots of the polynomial in the absolute value, which can be found numerically by using a suitable numerical algorithm.
Thus, with each mapping, the spectral coupling density is equipped with an additional pole.

\section{Benchmark of the reverse RC mapping}\label{APP:benchmark}

\begin{figure}
    \centering
    \includegraphics[width=0.45\textwidth,clip=true]{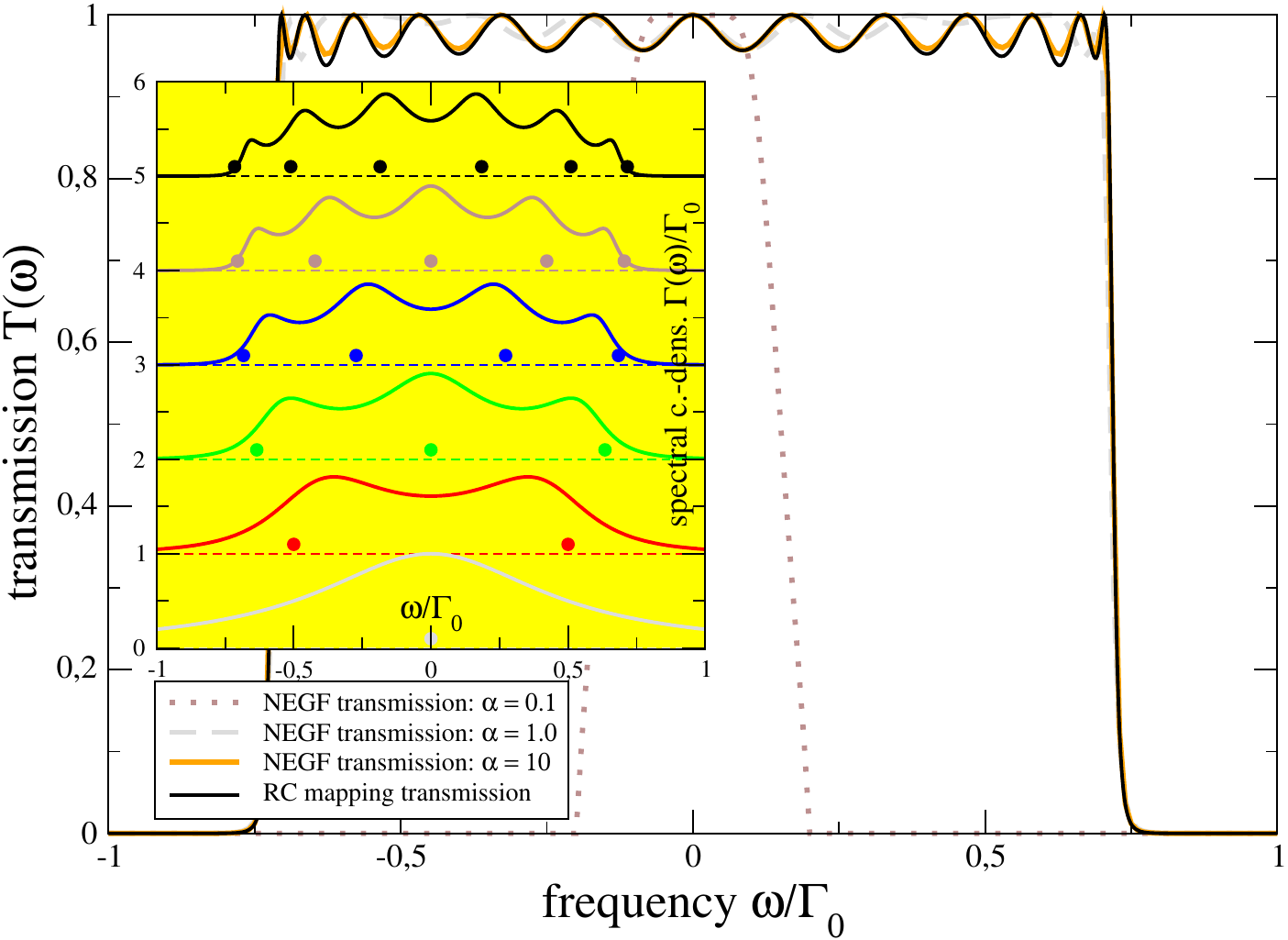}
    \caption{
    Main: Plot of the (symmetric) transmission function $T(\omega)$ through the optimized chain of 13 quantum dots as given in Tab.~\ref{TAB:optparam} and also shown in the inset of Fig.~\ref{FIG:uncert_entrop}, derived from the reverse mapping (black, see main text for explanations), and those derived for the same chain from a non-equilibrium Greens function approach (NEGF, dotted brown, dashed grey and solid orange) for spectral coupling densities shaped as~\eqref{EQ:sd_benchmark}.
    For flat external spectral coupling densities ($\alpha\to\infty$), they coincide.
    Inset: Plot of the corresponding mapped spectral coupling densities (shifted for clarity) after one reverse mapping (grey, bottom), two reverse mappings (red) and so on up to six reverse mappings (black, top) of a wideband reservoir. For the 13-dot chain discussed, the central quantum dot will then effectively see the black spectral coupling density for both reservoirs, yielding via Eq.~\eqref{EQ:transmission_formula} the black transmission in the main plot.
    } 
    \label{FIG:benchmark_revrcmapping}
\end{figure}

The transmission can alternatively be computed via the non-equilibrium Greens function technique~\cite{haug2008}, which requires the knowledge of a free Greens function.
In particular for reservoirs modeled by homogeneous tight-binding chains characterized by on-site energy $E$ and hopping amplitude $T$, one may choose the homogeneous tight-binding chain Greens function as the free one, which is well-known~\cite{economou2006}.
If the coupling between system and such a reservoir is described by tunnel amplitude $\tau$, such reservoirs lead to spectral coupling densities of semicircular (sc) form~\cite{boehling2018a}
\begin{align}\label{EQ:sd_benchmark}
    \Gamma_{\rm sc}(\omega) = \frac{2\tau^2}{T} \sqrt{1-\frac{(\omega-E)^2}{4 T^2}} \Theta(4T^2-(\omega-E)^2)\,,
\end{align}
which obviously are not of wideband shape.
However, sending both $T\to\infty$ and $\tau\to\infty$ while keeping $\frac{2\tau^2}{T} \equiv \Gamma_0$ constant, which can be achieved by scaling
$\tau_L=\tau_R=\sqrt{\alpha/2} \Gamma_0 = \tau$ and $T=\alpha\Gamma_0$ with dimensionless parameter $\alpha$,
we see that the wideband limit can be approached also by semicircular spectral coupling densities in the limit $\alpha\to\infty$.
Consequently, the chain transmission derived from the non-equilibrium Greens function formalism must converge to the transmission derived from the reverse reaction-coordinate mapping and Eq.~\eqref{EQ:transmission_formula} in this limit.
This is precisely what is seen in Fig.~\ref{FIG:benchmark_revrcmapping}.
For very flat reservoirs (large $\alpha$), the non-equilibrium Greens function transmission (orange) nearly fully agrees with the transmission obtained from the reverse mapping (black), and by further increasing $\alpha$ the curves would fully coincide (not shown).
This agreement for near-wideband reservoirs also indicates that our findings should also apply to more general initial reservoirs, as long as their bandwidth is large compared to the energy scales of the system.

The inset displays the sequence of mapped spectral coupling densities for either left or right wideband reservoir. 
After the first mapping, the spectral coupling density is Lorentzian and with each mapping, it acquires an additional pole.
An alternative perspective on the transformation of the spectral densities is that the boundary dots, i.e., the outermost dot (grey), the two outermost dots (red) and so on up to the six outermost dots (black) screen the remaining system from the wideband reservoir, leading to the observed structured spectral densities, compare also the bottom part of Fig.~\ref{FIG:quantum_dot_chain}.
As the parameters in Tab.~\ref{TAB:optparam} all scale with $\Gamma_0$, the system-reservoir interaction is not just a small perturbation to the Hamiltonian of the boundary dots, and therefore the peaks of the spectral densities do not fully coincide with the single-particle spectrum of the boundary dots (symbols of like color in inset).

\section{Equilibrium limit of the TUR}\label{APP:equilib}

The equilibrium limit can be recovered from the isothermal case $\beta=\beta_L=\beta_R$ in the limit $V\to 0$.
Writing the Fermi functions as $f_L(\omega)=[e^{\beta(\omega-(\bar\mu+V/2))}+1]^{-1}$ and
$f_R(\omega)=[e^{\beta(\omega-(\bar\mu-V/2))}+1]^{-1}$ with average chemical potential $\bar\mu$, we obtain from Eq.~\eqref{EQ:current_noise}
\begin{align}
    \lim_{V\to 0} S_M &= 2  \int T(\omega)
    \frac{e^{\beta (\omega+\bar\mu)}}{\left(e^{\beta\omega}+e^{\beta\bar\mu}\right)^2} \frac{d\omega}{2\pi}\,,\\
    \lim_{V\to 0} \frac{I_M}{\beta V} &= 
    \lim_{V\to 0} \frac{1}{\beta} \frac{dI_M}{dV} = \int T(\omega) 
    \frac{e^{\beta (\omega+\bar\mu)}}{\left(e^{\beta\omega}+e^{\beta\bar\mu}\right)^2} \frac{d\omega}{2\pi}\,,\nonumber
\end{align}
where we used l'Hospitals rule in the second line.
From this it follows that in equilibrium the STUR relation~\eqref{eq:stur} reaches equality $\lim_{V\to 0} \beta V \frac{S_M}{I_M} = 2$, regardless of the particular form of the transmission, as visible in Fig.~\ref{FIG:uncert_entrop}.
This just reflects the Johnson-Nyquist (fluctuation-dissipation) relation for the current~\cite{blanter2000a}
\begin{align}
    \lim_{V\to 0} S_M = \frac{2}{\beta} \lim_{V\to 0} \frac{dI_M}{dV}\,.
\end{align}

\section{Cooling and heating performance with ideal rectangular transmission functions}

\subsection{Cooling}\label{SEC:cooling_rect}

It has been noted before that rectangular transmission functions can be beneficial for cooling applications~\cite{whitney2014a}.
The optimal cooling performance is reached when the transmission is maximal where the integrand in the heat current is positive
\begin{align}
J_{L,1} = \frac{1}{2\pi} \int_{\mu_L}^{\bar\omega} (\omega-\mu_L) [f_L(\omega)-f_R(\omega)] d\omega\,.
\end{align}
When furthermore $\mu_R \ll \mu_L$ such that $f_R(\omega)\to 0$ in the integration interval and $\bar\omega\to\infty$, we obtain the upper bound
\begin{align}\label{EQ:maxcurrent}
J_{L,1} \le \frac{\pi}{24} T_L^2\,,
\end{align}
which after inserting appropriate units via $J_{L,1}\to\hbar J_{L,1}^{\rm SI}$ and $T_L \to k_{\rm B} T_L^{\rm SI}$ is just half~\cite{whitney2014a} Pendrys quantum bound~\cite{pendry1983a}.

The corresponding coefficient of performance is obtained by dividing the cooling current by the chemical work invested
\begin{align}
\kappa = \frac{I_E - \mu_L I_M}{(\mu_L-\mu_R) I_M} \Theta(I_E - \mu_L I_M) \le \frac{T_L}{T_R-T_L}\,,
\end{align}
where the Carnot bound can be generally seen from the positivity of the entropy production rate~\cite{nenciu2007a} and we use the Heaviside-$\Theta$ function to mind the range of applicability.
For the considered ideal limit (compare the blue transmission bar in Fig.~\ref{FIG:heatengine_refrigerator}) we can simplify this as
\begin{align}
\kappa &= \frac{\int_{\mu_L}^{\bar\omega} (\omega-\mu_L) [f_L(\omega)-f_R(\omega)] d\omega}{\int_{\mu_L}^{\bar\omega} (\mu_L-\mu_R) [f_L(\omega)-f_R(\omega)] d\omega}\\
&= \frac{\int_{\mu_L}^{\bar\omega} (\bar\omega-\mu_L) [f_L(\omega)-f_R(\omega)] d\omega}{\int_{\mu_L}^{\bar\omega} (\mu_L-\mu_R) [f_L(\omega)-f_R(\omega)] d\omega}\nn
&\qquad-\frac{\int_{\mu_L}^{\bar\omega} (\bar\omega-\omega) [f_L(\omega)-f_R(\omega)] d\omega}{\int_{\mu_L}^{\bar\omega} (\mu_L-\mu_R) [f_L(\omega)-f_R(\omega)] d\omega}\nn
&= \frac{\beta_R}{\beta_L-\beta_R} -\frac{\int_{\mu_L}^{\bar\omega} (\bar\omega-\omega) [f_L(\omega)-f_R(\omega)] d\omega}{\int_{\mu_L}^{\bar\omega} (\mu_L-\mu_R) [f_L(\omega)-f_R(\omega)] d\omega}\,.\nonumber
\end{align}
The first term in the last line implements the upper bound by the Carnot limit $\kappa_{\rm Ca} = \frac{T_L}{T_R-T_L}$, which is attained when the second term (which is detrimental to the coefficient of performance) is negligible.
We remark that analytic but lengthy results may be obtained and omit their explicit discussion here.

\subsection{Heat engine}\label{SEC:heatengine_rect}

The chemical work (electric power) extracted by the heat engine is generally given by
\begin{align}
P &= -I_M (\mu_L-\mu_R)\nn
&= \frac{1}{2\pi} \int (\mu_L-\mu_R) T(\omega) [f_R(\omega)-f_L(\omega)] d\omega\,.
\end{align}
Since $T(\omega)\ge 0$ and by assumption $\mu_L>\mu_R$, the power is maximized (for given thermal and potential biases) when the transmission is maximal $T(\omega) \to 1$ in the region where $f_L(\omega)<f_R(\omega)$, i.e., for $\omega>\bar\omega$, and vanishes elsewhere.
In regions where the power is positive, the energy to maintain it comes in as heat from the hot (right) reservoir
$J_R = -(I_E - \mu_R I_M)$, such that we can generally write for the efficiency (the Heaviside-$\Theta$ function is again merely used to mind the range of applicability)
\begin{align}
\eta = \frac{(\mu_L-\mu_R) I_M}{I_E - \mu_R I_M} \Theta(-(\mu_L-\mu_R) I_M) \le 1 - \frac{T_L}{T_R}\,,
\end{align}
and again the Carnot bound can be seen from the positivity of the entropy production rate~\cite{nenciu2007a}.
Inserting a rectangular transmission with ideal bounds, we specify
\begin{align}
\eta &= \frac{\int (\mu_L-\mu_R) T(\omega) [f_R(\omega)-f_L(\omega)] d\omega}{\int (\omega-\mu_R) T(\omega) [f_R(\omega)-f_L(\omega)] d\omega}\nn
&= \frac{1}{1+\frac{\int (\omega-\mu_L) T(\omega) [f_R(\omega)-f_L(\omega)] d\omega}{\int (\mu_L-\mu_R) T(\omega) [f_R(\omega)-f_L(\omega)] d\omega}}\,,
\end{align}
where we have used $\omega-\mu_R=(\omega-\mu_L)+(\mu_L-\mu_R)$ in the denominator of the first line to split the integral.
One can see that Carnot efficiency $\eta_{\rm Ca} = 1 - \frac{\beta_R}{\beta_L}$ can be approached for a very narrow transmission with effectively zero power output.
Therefore, we rather focus on the efficiency at maximum power, where we first maximize the power with respect to position and width of the transmission window
\begin{align}\label{EQ:maxpower}
P_1 &= \frac{(\mu_L-\mu_R)}{2\pi} \int_{\bar\omega}^\infty [f_R(\omega)-f_L(\omega)] d\omega\nn
&= \frac{(\mu_L-\mu_R)}{2\pi} \int_0^\infty [f_R(\omega+\bar\omega)-f_L(\omega+\bar\omega)] d\omega\nn
&= \frac{1}{2\pi} \left(\frac{\beta_L-\beta_R}{\beta_L \beta_R}\right)^2 x \ln \left(1+e^{-x}\right)
\;:\;
x\equiv \frac{\beta_L \beta_R}{\beta_L-\beta_R}V\nn
&\le \frac{1}{2\pi} \left(\frac{\beta_L-\beta_R}{\beta_L \beta_R}\right)^2 (0.31635)\,,
\end{align}
where the last bound is obtained numerically by maximizing with respect to the voltage $V=\mu_L-\mu_R$, which is saturated at $x\approx 1.14455$.
In the proper units ($P\to P_{\rm SI} \hbar$ and $T_\nu \to k_{\rm B} T_\nu^{\rm SI})$ this reproduces the bound in Ref.~\cite{whitney2014a}). 
Therefore, the efficiency for optimal transmission becomes
\begin{align}
\eta_1 &= \frac{1}{1+\frac{\int\limits_{\bar\omega}^\infty (\omega-\mu_L) [f_R(\omega)-f_L(\omega)] d\omega}{\int\limits_{\bar\omega}^\infty (\mu_L-\mu_R) [f_R(\omega)-f_L(\omega)] d\omega}}\nn
&= \frac{1}{1+\frac{\int\limits_{0}^\infty (\omega+\bar\omega-\mu_L) [f_R(\omega+\bar\omega)-f_L(\omega+\bar\omega)] d\omega}{\int\limits_{0}^\infty (\mu_L-\mu_R) [f_R(\omega+\bar\omega)-f_L(\omega+\bar\omega)] d\omega}}\nn
&= \frac{1}{1+\frac{\beta_R}{\beta_L-\beta_R} + \frac{\int\limits_{0}^\infty \omega [f_R(\omega+\bar\omega)-f_L(\omega+\bar\omega)] d\omega}{\int\limits_{0}^\infty (\mu_L-\mu_R) [f_R(\omega+\bar\omega)-f_L(\omega+\bar\omega)] d\omega}}\nn
&= \frac{1}{1+\frac{\beta_R}{\beta_L-\beta_R} + \frac{\beta_L+\beta_R}{\beta_L - \beta_R} \frac{[-{\rm Li}_2(-e^{-x})]}{x \ln[1+e^{-x}]}}\nn
&= \eta_{\rm Ca} \frac{1}{1+\left(1+\frac{\beta_R}{\beta_L}\right) \frac{[-{\rm Li}_2(-e^{-x})]}{x \ln[1+e^{-x}]}}
\,,
\end{align}
where ${\rm Li}_2(y)$ denotes the dilogarithm and $x$ is defined as before.
At the voltage that maximizes the power, we thus have $\eta_1 = \frac{\eta_{\rm Ca}}{1+0.93593 \left(1+\frac{\beta_R}{\beta_L}\right)}$, which is well below the Chambadal~\cite{chambadal1957}-Novikov~\cite{novikov1958a}-Curzon-Ahlborn~\cite{curzon1975a} efficiency as expected at maximum power~\cite{vandenbroeck2005a}.

\end{document}